\documentclass[aps,prc,showpacs,superscriptaddress]{revtex4}
\usepackage{multirow}
\usepackage{graphicx}
\usepackage{amsmath}

\newcommand{\sqrtsNN}{\mbox{$\sqrt{\mathrm{s}_{_{\mathrm{NN}}}}$}}
\newcommand{\axi}{$\overline{\Xi}^+$ }
\newcommand{\xim}{$\Xi^-$ }
\newcommand{\alam}{$\overline{\Lambda}$ }
\newcommand{\lam}{$\Lambda$ }
\newcommand{\ks}{$\mathrm{K}^{0}_{S}$ }
\newcommand{\omm}{$\Omega^-$ }
\newcommand{\aom}{$\overline{\Omega}^+$ }

\newcommand{\pt}{$p_T$ }
\newcommand{\et}{$\eta$ }
\def \GeVc {\mbox{$\mathrm{GeV}/c$}}
\def \lt {\mbox{$\ <\ $}}
\def \gt {\mbox{$\ >\ $}}

\def \auau  {$Au + Au$ }

\def \etal   {\mbox{$\mathrm{\it et\ al.}$}}
\newcommand{\mean}[1]{\left\langle #1 \right\rangle}
\usepackage{color}

\begin{document}

\title{Centrality dependence of charged hadron and strange hadron
elliptic flow \\
from \sqrtsNN~= 200 GeV \auau collisions}

\affiliation{Argonne National Laboratory, Argonne, Illinois 60439}
\affiliation{University of Birmingham, Birmingham, United Kingdom}
\affiliation{Brookhaven National Laboratory, Upton, New York 11973}
\affiliation{California Institute of Technology, Pasadena, California 91125}
\affiliation{University of California, Berkeley, California 94720}
\affiliation{University of California, Davis, California 95616}
\affiliation{University of California, Los Angeles, California 90095}
\affiliation{Universidade Estadual de Campinas, Sao Paulo, Brazil}
\affiliation{Carnegie Mellon University, Pittsburgh, Pennsylvania 15213}
\affiliation{University of Illinois at Chicago, Chicago, Illinois 60607}
\affiliation{Creighton University, Omaha, Nebraska 68178}
\affiliation{Nuclear Physics Institute AS CR, 250 68 \v{R}e\v{z}/Prague, Czech Republic}
\affiliation{Laboratory for High Energy (JINR), Dubna, Russia}
\affiliation{Particle Physics Laboratory (JINR), Dubna, Russia}
\affiliation{University of Frankfurt, Frankfurt, Germany}
\affiliation{Institute of Physics, Bhubaneswar 751005, India}
\affiliation{Indian Institute of Technology, Mumbai, India}
\affiliation{Indiana University, Bloomington, Indiana 47408}
\affiliation{Institut de Recherches Subatomiques, Strasbourg, France}
\affiliation{University of Jammu, Jammu 180001, India}
\affiliation{Kent State University, Kent, Ohio 44242}
\affiliation{University of Kentucky, Lexington, Kentucky, 40506-0055}
\affiliation{Institute of Modern Physics, Lanzhou, China}
\affiliation{Lawrence Berkeley National Laboratory, Berkeley, California 94720}
\affiliation{Massachusetts Institute of Technology, Cambridge, MA 02139-4307}
\affiliation{Max-Planck-Institut f\"ur Physik, Munich, Germany}
\affiliation{Michigan State University, East Lansing, Michigan 48824}
\affiliation{Moscow Engineering Physics Institute, Moscow Russia}
\affiliation{City College of New York, New York City, New York 10031}
\affiliation{NIKHEF and Utrecht University, Amsterdam, The Netherlands}
\affiliation{Ohio State University, Columbus, Ohio 43210}
\affiliation{Panjab University, Chandigarh 160014, India}
\affiliation{Pennsylvania State University, University Park, Pennsylvania 16802}
\affiliation{Institute of High Energy Physics, Protvino, Russia}
\affiliation{Purdue University, West Lafayette, Indiana 47907}
\affiliation{Pusan National University, Pusan, Republic of Korea}
\affiliation{University of Rajasthan, Jaipur 302004, India}
\affiliation{Rice University, Houston, Texas 77251}
\affiliation{Universidade de Sao Paulo, Sao Paulo, Brazil}
\affiliation{University of Science \& Technology of China, Hefei 230026, China}
\affiliation{Shanghai Institute of Applied Physics, Shanghai 201800, China}
\affiliation{SUBATECH, Nantes, France}
\affiliation{Texas A\&M University, College Station, Texas 77843}
\affiliation{University of Texas, Austin, Texas 78712}
\affiliation{Tsinghua University, Beijing 100084, China}
\affiliation{Valparaiso University, Valparaiso, Indiana 46383}
\affiliation{Variable Energy Cyclotron Centre, Kolkata 700064, India}
\affiliation{Warsaw University of Technology, Warsaw, Poland}
\affiliation{University of Washington, Seattle, Washington 98195}
\affiliation{Wayne State University, Detroit, Michigan 48201}
\affiliation{Institute of Particle Physics, CCNU (HZNU), Wuhan 430079, China}
\affiliation{Yale University, New Haven, Connecticut 06520}
\affiliation{University of Zagreb, Zagreb, HR-10002, Croatia}

\author{B.I.~Abelev}\affiliation{University of Illinois at Chicago, Chicago, Illinois 60607}
\author{M.M.~Aggarwal}\affiliation{Panjab University, Chandigarh 160014, India}
\author{Z.~Ahammed}\affiliation{Variable Energy Cyclotron Centre, Kolkata 700064, India}
\author{B.D.~Anderson}\affiliation{Kent State University, Kent, Ohio 44242}
\author{D.~Arkhipkin}\affiliation{Particle Physics Laboratory (JINR), Dubna, Russia}
\author{G.S.~Averichev}\affiliation{Laboratory for High Energy (JINR), Dubna, Russia}
\author{Y.~Bai}\affiliation{NIKHEF and Utrecht University, Amsterdam, The Netherlands}
\author{J.~Balewski}\affiliation{Massachusetts Institute of Technology, Cambridge, MA 02139-4307}
\author{O.~Barannikova}\affiliation{University of Illinois at Chicago, Chicago, Illinois 60607}
\author{L.S.~Barnby}\affiliation{University of Birmingham, Birmingham, United Kingdom}
% ...

\collaboration{STAR Collaboration}\noaffiliation




\date{\today}

\begin{abstract}
We present STAR results on the elliptic flow $v_2$ of charged
hadrons, strange and multi-strange particles from \sqrtsNN~= 200 GeV
\auau collisions at RHIC. The detailed study of the centrality
dependence of $v_2$ over a broad transverse momentum range is
presented. Comparison of different analysis methods are made in order
to estimate systematic uncertainties. In order to discuss the non-flow
effect, we have performed the first analysis of $v_2$ with the
Lee-Yang Zero method for \ks and \lam \!.

In the relatively low \pt region, \pt $\le$ 2 \GeVc, a scaling with
$m_T - m$ is observed for identified hadrons in each centrality bin
studied. However, we do not observe $v_2(p_T)$ scaled by the
participant eccentricity to be independent of centrality. At higher
\pt\!, 2 \GeVc\: $\le$ \pt $\le$ 6 \GeVc, $v_2$ scales with quark number
for all hadrons studied. For the multi-strange hadron $\Omega$, which
does not suffer appreciable hadronic interactions, the values of $v_2$
are consistent with both $m_T -m$ scaling at low \pt and
number-of-quark scaling at intermediate $p_T$. As a function of
collision centrality, an increase of $p_T$-integrated $v_2$ scaled by
the participant eccentricity has been observed, indicating a stronger
collective flow in more central \auau collisions.
\end{abstract}

\maketitle

\section{Introduction}
\label{intro}

The event azimuthal anisotropy with respect to the reaction plane has
been widely studied in order to evaluate the collective behavior of
the matter produced in high-energy nuclear
collisions~\cite{yingchao98,v2Methods,Borghini}. The initial
configuration space anisotropy is expected to be self-quenched by
expansion and reduced by frequent re-scatterings in the hot and dense
medium created in such collisions. The final observed momentum space
anisotropies, therefore, carry information about the early stage
collision dynamics~\cite{sorge97,ollitrault92,shuryak01}. The
experimental results of the second harmonic azimuthal anisotropy,
elliptic flow, $v_2$, from \auau collisions have demonstrated the
development of partonic
collectivity~\cite{PIDv2130,flowPRC,starv21,starklv2}. Further
detailed analyses of the hadron mass dependence of $v_2$ suggest that
the system has been in the deconfined state with constituent quark
degrees of freedom prior to
hadronization~\cite{starwp,msv2,mv03}. Furthermore, results of
multi-strange hadron transverse momentum distributions and $v_2$
indicate that the system reached thermalization at the partonic
stage~\cite{starompt,phenix_fv2,star_fv2,nagle}.

Hydrodynamic model calculations, with the assumption of ideal fluid
behavior (no viscosity), have been successful when compared with the
experimental data at RHIC~\cite{starwp,shuryak01,flowPRC,heinz04}. It
should be noted that up to now, the discussions of the underlying
dynamics of the thermalization at RHIC are inconclusive. Some initial
evidence for thermalization was provided by the quantitative agreement
of $v_2$ results between ideal hydrodynamic model calculations and
data for identified hadrons $\pi$, $K$, $p$ and
$\Lambda$~\cite{starwp,phenixwp} from {\it minimum bias} (0 - 80\%
centrality) \auau collisions~\cite{gyulass_riken03}. As shown in
Refs.~\cite{flowPRC,PIDv2130}, ideal hydrodynamic model calculations
have failed to reproduce the centrality dependence of $\pi$ and $p$
$v_2$ in \auau collisions. In addition, the discussion based on the
integrated $v_2 / \varepsilon_{part}$ of charged hadrons suggests
possible thermalization {\it only} for the most central collisions at
RHIC (see Refs.~\cite{voloshin06,STARcum} and references therein).
Here the participant eccentricity, $\varepsilon_{part}$, is the
initial configuration space eccentricity of the participants.  From
peripheral to the most central \auau collisions, the values of $v_2 /
\varepsilon_{part}$ increase as a function of the scaled charged
hadron multiplicity, as predicted by a model calculation assuming the
{\it low density limit}~\cite{ln} of single forward nucleon-nucleon
collisions. This analysis indicates that the system has probably not
reached thermalization for most peripheral \auau collisions.

Hydrodynamic model calculations predict a characteristic dependence of
the elliptic flow and transverse momentum spectra on particle mass and
collision centrality. Nevertheless the comparisons made so far have
been mostly restricted to identified hadrons from minimum bias
collisions or integrated $v_2$ of charged
hadrons~\cite{voloshin06}. Systematic comparisons for identified
hadrons at different collision centralities are still
scarce~\cite{flowPRC}. In order to fill this gap and further advance
our understanding of the properties of the medium created in
high-energy nuclear collisions, in this article we report the
centrality dependence of the azimuthal anisotropy parameter $v_2$
(elliptic flow) in \sqrtsNN~= 200 GeV \auau collisions. The centrality
dependence of $v_2$ for identified hadrons $K_S^0$, $\Lambda$, $\Xi$,
and $\Omega$, and the scaling properties as a function of number of
quarks within a given hadron and the transverse kinetic energy $m_T-m$
are reported. Results from the Lee-Yang Zero method~\cite{LYZ_PLB,LYZ}
for unidentified charged hadrons ($h^+ + h^-$), \ks and \lam are also
reported. In complex nuclear collisions different systematic errors on
$v_2$ can arise from different analysis methods. In this paper, the
systematic errors are analyzed by comparing the standard event plane
method~\cite{v2Methods,flowPRC} with results from the Lee-Yang
Zero~\cite{LYZ}, four-particle cumulant~\cite{STARcum,cum}, and $\eta$
subevent~\cite{flowPRC} methods.

The paper is organized in the following way: we discuss experimental
cuts, data selections and methods used for unidentified charged
hadrons and identified hadrons in
Section~\ref{methods}. Section~\ref{sect_result} gives the results on
$v_2$ for unidentified charged hadrons and identified hadrons along
with a discussion of the systematic errors extracted from different
analysis methods. Section~\ref{discus} presents a comparison with
model calculations as well as a discussion of scaling and the
systematics of the $v_2$($p_T$) distributions from SPS (\sqrtsNN =
17.3 GeV) and RHIC (\sqrtsNN = 62.4 and 200 GeV). Finally, the summary
of the analysis and the outlook are presented in Section~\ref{concl}.

\section{Methods and Analysis}
\label{methods}

In this paper, if $v_2$ is used without curly brackets it is an
abbreviation for $v_2\{EP_2\}$, that is, $v_2$ relative to the second
harmonic event plane. The systematic uncertainty from the event plane
resolution is constant at each centrality and expected to be smaller
than that of the observed differential flow, and is not folded
in. Other systematic errors are discussed in
Section~\ref{sect_result}.

\subsection{Data sets}

For this study, the high statistics data from \sqrtsNN~= 200 GeV
\auau collisions collected by the STAR experiment during RHIC's
fourth year (2004) of data taking were analyzed. STAR's main time
projection chamber (TPC)~\cite{STARtpc} was used for tracking and
identification of charged particles. The TPC records the hits used for
reconstructions of the tracks of the particles, enabling the
measurement of the momenta of the particles, and identifying the
particles by measuring their ionization energy loss.  The TPC provides
complete azimuthal coverage and complete tracking for charged
particles within $\pm 1.8$ units of pseudorapidity. The two forward
time projection chambers (FTPC) cover two sides of the collision with
$2.5 \lt \mid \eta \mid \lt 4.0.$ The FTPCs also provide a tool for
studying non-flow effects. About 25 million minimum bias events
(0--80\% most central of the hadronic interaction cross section) were
analyzed, which increased the statistics for flow analysis by more
than a factor of 10 compared to the previous
measurements~\cite{flowPRC,starklv2,msv2}.

The centrality definition of an event was based on the number of
charged tracks in the TPC with track quality cuts, which are $|\eta|$
$\lt$ 0.5, a distance of closest approach to the primary vertex (DCA)
less than 3 cm, and fit points more than 15.  These events were grouped
into three centrality bins, which were central (0--10\%), mid-central
(10--40\%), and peripheral (40--80\%). In addition, the central dataset
(0--10\%) was enhanced by online triggering on the most central events
with the Zero-Degree Calorimeter (ZDC)~\cite{ZDC}, thereby getting an
additional $\sim$19 million events for a similar centrality
bin. Within the statistical uncertainty, the results from the central
trigger dataset were consistent with those from the minimum bias
trigger.

The analyzed charged particles were identified as the track helix in
the TPC magnetic field. The charged tracks were selected with a
transverse momentum range of $0.15 \lt p_T \lt 2.0$ \GeVc\: unless
indicated otherwise, and a pseudorapidity range of $\mid \eta \mid
\lt 1.0$. For the Lee-Yang Zero product generating function analysis
the \et interval increased to $\mid \eta \mid \lt 1.3$ in order to
obtain more particles. A minimum of 15 track fit hits and a ratio of
hits to maximum possible hits $\gt 0.52$ was also required. To improve
the selection of good tracks from the primary collisions, the distance
of closest approach of the analyzed tracks to the event vertex had to
be less than 2 cm. Tracks of charged daughter particles stemming from
weak decay, which tend to be at large distances, are not subject to
this cut.

\subsection{Particle identification}

We identified \ks, \lam ($\overline{\Lambda}$), $\Xi^{-}$
($\overline{\Xi}^{+}$), $\Omega^{-}$ ($\overline{\Omega}^{+}$) through
their decay channel: \ks $\rightarrow \pi^{+} + \pi^{-}$, \lam
$\rightarrow p + \pi^{-}$ ($\overline{\Lambda} \rightarrow
\overline{p} + \pi^{+}$), $\Xi^{-} \rightarrow$ \lam $+\ \pi^{-}$
($\overline{\Xi}^{+} \rightarrow$ $\overline{\Lambda}$+\ $\pi^{+}$),
$\Omega^{-}\rightarrow$ \lam $+\ K^{-}$
($\overline{\Omega}^{+}\rightarrow$ $\overline{\Lambda} +\ K^{+}$).
The charged pions, kaons and protons were identified via their energy
loss in the TPC~\cite{STARtpc}.  According to the (multi)strange
particle decay properties, topological cuts and kinematic cuts were
applied to reduce the combinatorial background. The detailed
description of the analysis method can be found in
Refs~\cite{klv2_130GeV,starklv2} for \ks and \lam, and
Refs.~\cite{msp_130GeV,msv2} for $\Xi$ and $\Omega$.

\begin{figure}[ht]
\includegraphics[width=0.7\textwidth]{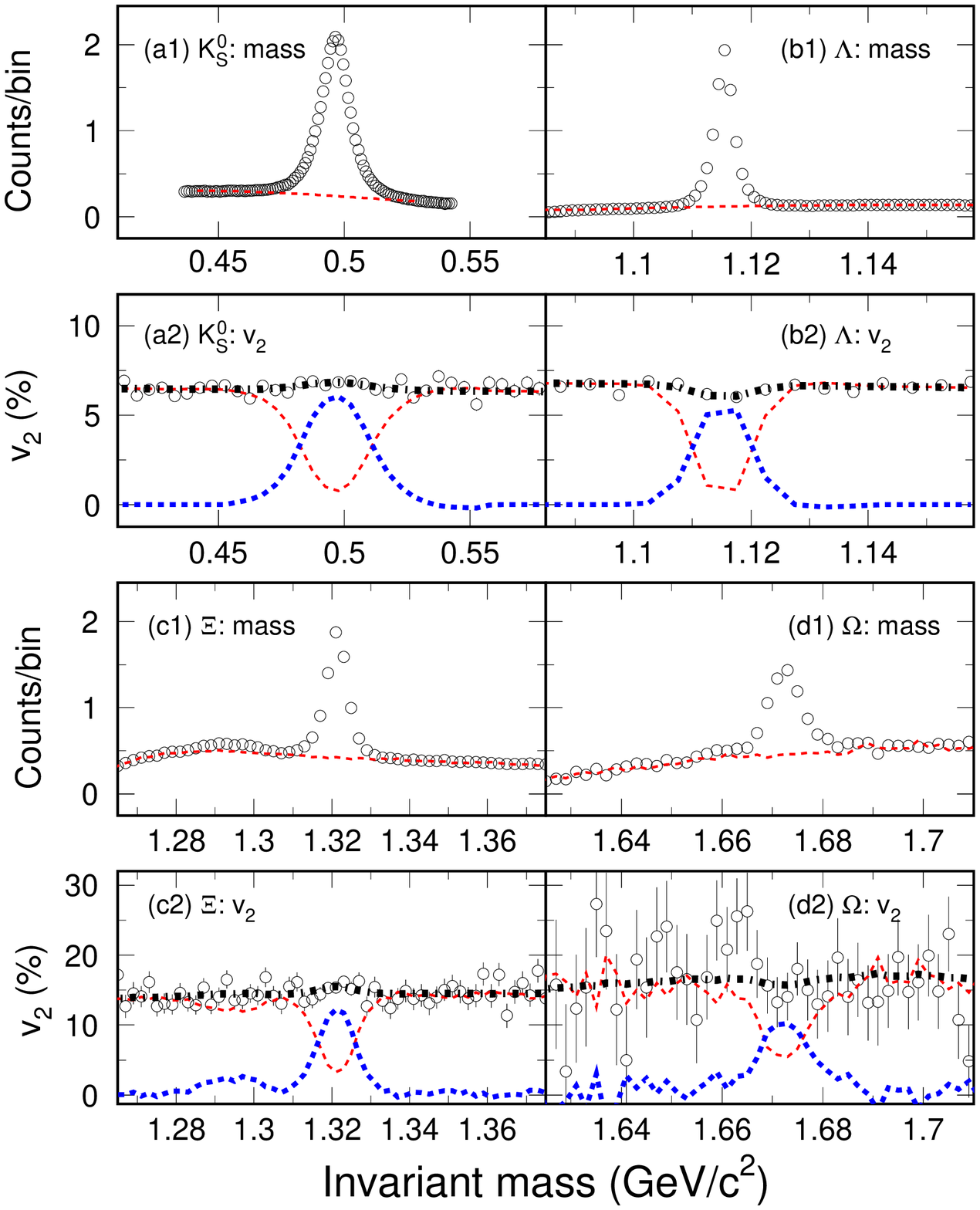}
\caption{(Color online) Plots (a1)--(d1) represent the invariant
mass distributions for \ks (1.4 $\le p_T \le$ 1.6 \GeVc), \lam (1.4
$\le p_T \le$ 1.6 \GeVc), $\Xi$ (2.3 $\le p_T \le$ 2.6 \GeVc), and
$\Omega$ (2.5 $\le p_T \le$ 3.0 \GeVc), respectively, from \sqrtsNN
= 200 GeV minimum bias (0--80\%) \auau collisions. The dashed lines
are the background distributions. The corresponding data for the
$v_2$ distributions are shown in plots (a2)--(d2) as open circles.
The thick-dashed, thin-dashed and the dot-dashed lines represent the
relative contributions of $v_2(Sig)$, $v_2(Bg)$, and $v_2(Sig+Bg)$,
respectively. For clarity, the invariant mass plots for \ks\!\!,
\lam\!\!, $\Xi$, and $\Omega$, are scaled by 1/50000, 1/170000,
1/2.5, and 1/3, respectively. The error bars are shown only for the
statistical uncertainties.} \label{v2vsm}
\end{figure}

Figure~\ref{v2vsm} shows the invariant mass distributions for (a1)
\ks, (b1) \lam + \alam, (c1) \xim + \axi and (d1) \omm + \aom for a
given \pt bin from \sqrtsNN~= 200 GeV minimum bias (0--80\%) \auau
collisions. Clear signal peaks are seen at the values expected for
the particle mass above the combinatorial background. The measured
invariant mass distributions contain both signal ($Sig$) and
combinatorial background ($Bg$).  For \ks and \lam, the measured
invariant mass distributions were fitted by a polynomial (up to
4$^{th}$ order), which represents the background, and a
double-Gaussian function, which represents the signal. The
double-Gaussian was used because of tails on the distribution. For
multi-strange baryons $\Xi$ and $\Omega$, the Bg was estimated by
rotating the transverse momentum of the daughter \lam by
$180^{\circ}$. This operation breaks the correlation between the
$\Lambda$ and the other daughter particle. The resulting invariant
mass distributions provide a good approximation for the true
background distribution. The detailed description of the method can
be found in Ref.~\cite{msv2}.

For $v_2$ of the identified particles, \ks, \lam + \alam, \xim +
\axi and \omm + \aom, the  $v_2$ versus $m_{inv}$ method is used in
this analysis~\cite{v2minv}. Since $v_2$ is additive, one can write
the total $v_{2}^{Sig+Bg}$ as a sum of $Sig$ and $Bg$ contributions
weighted by their relative yields:

\begin{equation} \label{v2minv} v_{2}^{Sig+Bg}(m_{inv}) =
  v_{2}^{Sig}\cdot\frac{Sig}{Sig+Bg}(m_{inv})\ +\
  v_{2}^{Bg}(m_{inv})\cdot\frac{Bg}{Sig+Bg}(m_{inv}) .
\end{equation}

This method involves the calculation of $v_{2}^{Sig+Bg}$ as a function
of $m_{inv}$ and then fitting the distribution using
Eq.~(\ref{v2minv}) with measured relative yields and parameterizations
of $v_{2}^{Sig}$ and $v_{2}^{Bg}(m_{inv})$. The
$\frac{Bg}{Sig+Bg}(m_{inv})$ distribution is the $Bg$ divided by
($Sig+Bg$). The $\frac{Sig}{Sig+Bg}(m_{inv})$ distribution is simply
calculated by $1\ -\ \frac{Bg}{Sig+Bg}(m_{inv})$. The term
$v_{2}^{Bg}(m_{inv})$ is parameterized as a linear function in order
to take care of the non-constant $v_{2}^{Bg}$ value as a function of
$m_{inv}$. The fit result $v_{2}^{Sig}$ is the final observed $v_2$.
The fit results for \ks, \lam + \alam, \xim + \axi and \omm + \aom are
shown, as dot-dashed lines, in Fig.~\ref{v2vsm} (a2), (b2), (c2), and
(d2), respectively. Note that the anisotropy varies as a function of
\pt and hadron mass. In this figure, the $v_2$ are shown for different
hadrons with the \pt cuts that are listed in the caption. How this
method works well for measuring signal $v_2$ is explained as
following: a set of data points are used in the fit over a wide
$m_{inv}$ region for $Sig$ and $Bg$. Data points far from the mass
peak constrain $v_{2}^{Bg}(m_{inv})$ since pure $Bg$ is expected in
this region. (The disagreement at $M_{inv} \sim 1.29\: \GeVc^2\:$ in
(c1) is caused by mis-identified hadrons in the Lambda-pion
combinations, which is explained in Ref.~\cite{msv2}.) Under the peak,
the $v_{2}^{Sig+Bg}(m_{inv})$ is dominated by the $Sig$
distribution. Finally, the $v_2$ signal is extracted by the fitting
method shown in Eq.~\ref{v2minv}.

The results obtained with this technique are in good agreement with
the ones from the method used previously~\cite{v2Methods}. Note that
the subtraction procedure used to extract the $v_2$ signal for a
given identified particle is independent of the flow correlations.
The $v_2$ distributions of the overall signal and background are
evaluated by a specific flow analysis method. These methods will be
discussed in Section~\ref{subsect_flowmethod}.

\subsection{Flow analysis methods}
\label{subsect_flowmethod}

The systematic uncertainty of the Event Plane method is evaluated by
comparing the results to those obtained by other techniques for
measuring anisotropic flow. The various methods have different
sensitivities to non-flow effects and $v_2$ fluctuations, and such
studies provide information on the magnitude of the systematic
uncertainty. Non-flow effects are correlations not associated with the
reaction plane, and include resonance decays, HBT correlations, final
state interactions, and jets, to the extent that they do not
participate in the flow.

\subsubsection{Event Plane method}

The essence of the Event Plane method~\cite{v2Methods} is to first
estimate the reaction plane. The estimated reaction
plane is called the event plane and is determined by the anisotropic
flow itself for each harmonic of the Fourier expansion of the
anisotropic flow. The event flow vector $Q_2$ and the event plane
angle $\Psi_2$ are defined by the following equations:
\begin{equation} \label{Q2x} Q_2\cos(2\Psi_2)\ =\ Q_{2x}\ = \
\sum_{i}w_i\cos(2\phi_i)
\end{equation}
\begin{equation} \label{Q2y} Q_2\sin(2\Psi_2)\ =\ Q_{2y}\ = \
\sum_{i}w_i\sin(2\phi_i)
\end{equation}
\begin{equation} \label{Psi} \Psi_2\ =\
\left(\tan^{-1}\frac{Q_{2y}}{Q_{2x}}\right)/2
\end{equation}
where the sum goes over all the particles $i$ used in the event plane
calculation. $\phi_i$ and $w_i$ are the lab azimuthal angle and the
weight for the particle $i$, respectively. In this analysis, the
weights are taken to be the value of \pt in \GeVc\: up to 2 \GeVc\:
and then constant at 2.0 above that \pt.

The observed $v_{2}$ is the second harmonic of the azimuthal
distribution of particles with respect to this event plane:
\begin{equation} \label{v2obs} v_{2}^{obs}\ =\ \langle
\cos[2(\phi-\Psi_2)]\rangle
\end{equation}
where angle brackets denote an average over all particles with their
azimuthal angle $\phi$ in a given phase space. Since finite
multiplicity limits the resolution in estimating the angle of the
reaction plane, the real $v_2$ has to be corrected for the event plane
resolution by
\begin{equation} \label{v2EP2} v_{2}\ =\
\frac{v_2^{obs}}{\langle \cos[2(\Psi_2-\Psi_r)]\rangle}
\end{equation}
where brackets denote an average over a large event sample, and
$\Psi_r$ is the angle of the reaction plane. The event plane
resolution is estimated by the correlation of the event planes of two
subevents. The event plane resolution for the subevents with the
assumption of pure flow correlations between the subevents is
\begin{equation}
\langle \cos[2(\Psi_{2}^{A}-\Psi_{r})] \rangle = \sqrt { \langle
\cos[2(\Psi_{2}^{A}-\Psi_{2}^{B})] \rangle } \label{subEPres}
\end{equation}
where A and B denote two subgroups of tracks. In this analysis, we use
two random subevents with equal numbers of particles. Further, the
full event plane resolution is obtained from the resolution of the
subevents:
\begin{equation} \langle
\cos[2(\Psi_{2}-\Psi_{r})] \rangle = C \langle
\cos[2(\Psi_{2}^{A}-\Psi_{r})] \rangle \label{EPres}
\end{equation}
where C is a constant calculated from the known multiplicity
dependence of the resolution~\cite{v2Methods}. In the case of low
resolution ($\lt 0.5$), C is equal to $\sqrt2$~\cite{v2Methods}.  The
actual event plane resolutions, for the centrality bins used in this
analysis of 0-10\%, 10-40\% and 40-80\%, were 0.658 $\pm$ 0.0006,
0.818 $\pm$ 0.0002 and 0.694 $\pm$ 0.0004, respectively.

\subsubsection{$\eta$-subevent method}
\label{etaSubs}

The $\eta$-subevent method attempts to reduce the contribution from
non-flow effects (mostly due to short range correlations) by
correlating particles separated in pseudorapidity. This technique is
similar to the event plane method, except one defines the event flow
vector for each particle based on particles measured in the opposite
hemisphere in pseudorapidity:
\begin{equation}
\label{etaSubeventMethod}
  v_{2}\{\eta_\pm\} = \frac{\left<\cos [2(\phi_{\eta_\pm}-\Psi_{2,
  \eta_\mp})]\right>}{\sqrt{\left< \cos [2 (\Psi_{2, \eta_+}-\Psi_{2, \eta_-})]\right>}}.
\end{equation}
Here $\Psi_{\rm 2, \eta_+}$ ($\Psi_{2, \eta_-}$) is the second
harmonic event plane angle defined for particles with positive
(negative) pseudorapidity. An $\eta$ gap of $\mid \eta \mid \lt 0.075$
between positive and negative pseudorapidity subevents is introduced
in order to guarantee that non-flow effects are reduced by enlarging
the separation between the correlated particles. In
Eq.~(\ref{etaSubeventMethod}) the non-flow effects (correlations) are
reduced in both the observed flow (numerator) and the event plane
resolution (denominator). Depending on the nature of the remaining
non-flow effects, $v_2$ measured this way may have values which are
either lower or higher than those obtained with the standard method.

\subsubsection{4-particle cumulant method}

A method to calculate $v_2$ from true four-particle correlations was
developed in Ref.~\cite{cum} and it has already been used by
STAR~\cite{STARcum}. It uses the cumulant relation
\begin{equation}
\label{cumulant}
  C\{4\} \equiv \mean{u_{n,1} u_{n,2} u_{n,3}^* u_{n,4}^*}
  -2 \mean{u_{n,1} u_{n,2}^*}^2
  = -v_n^4\{4\} \,,
\end{equation}
where $u_{n,j}=e^{in\phi_j}$. The cumulant allows one to subtract the
two-particle correlations, including two-particle non-flow, from the
four-particle correlations. In practice, cumulants are calculated
using the generating function from Ref.~\cite{cum} and described in
Ref.~\cite{STARcum}:
\begin{equation}
  G_n(z)  =  \prod_{j=1}^M
  \left( 1 + \frac{z^* u_{n,j} + z u_{n,j}^*}{M} \right),
\label{eq:newG0}
\end{equation}
where $z \equiv |z| e^{i\alpha}$ is an arbitrary complex number, with
$z^*$ denoting its complex conjugate. The cumulants are related to the
generating function by
\begin{equation}
  M \cdot \left( \mean{ G_n(z)}^{1/M} - 1 \right) =
  \sum_{k} \frac{|z|^{2k}}{(k!)^2} C\{2k\} \,.
\label{eq:defc2}
\end{equation}
The fit to the $C\{4\}$ term is needed. The fourth root of the
negative of it gives $v_2\{4\}$.

\subsubsection{Lee-Yang Zero method}

\begin{figure}[ht]
\includegraphics[width=0.4\textwidth]{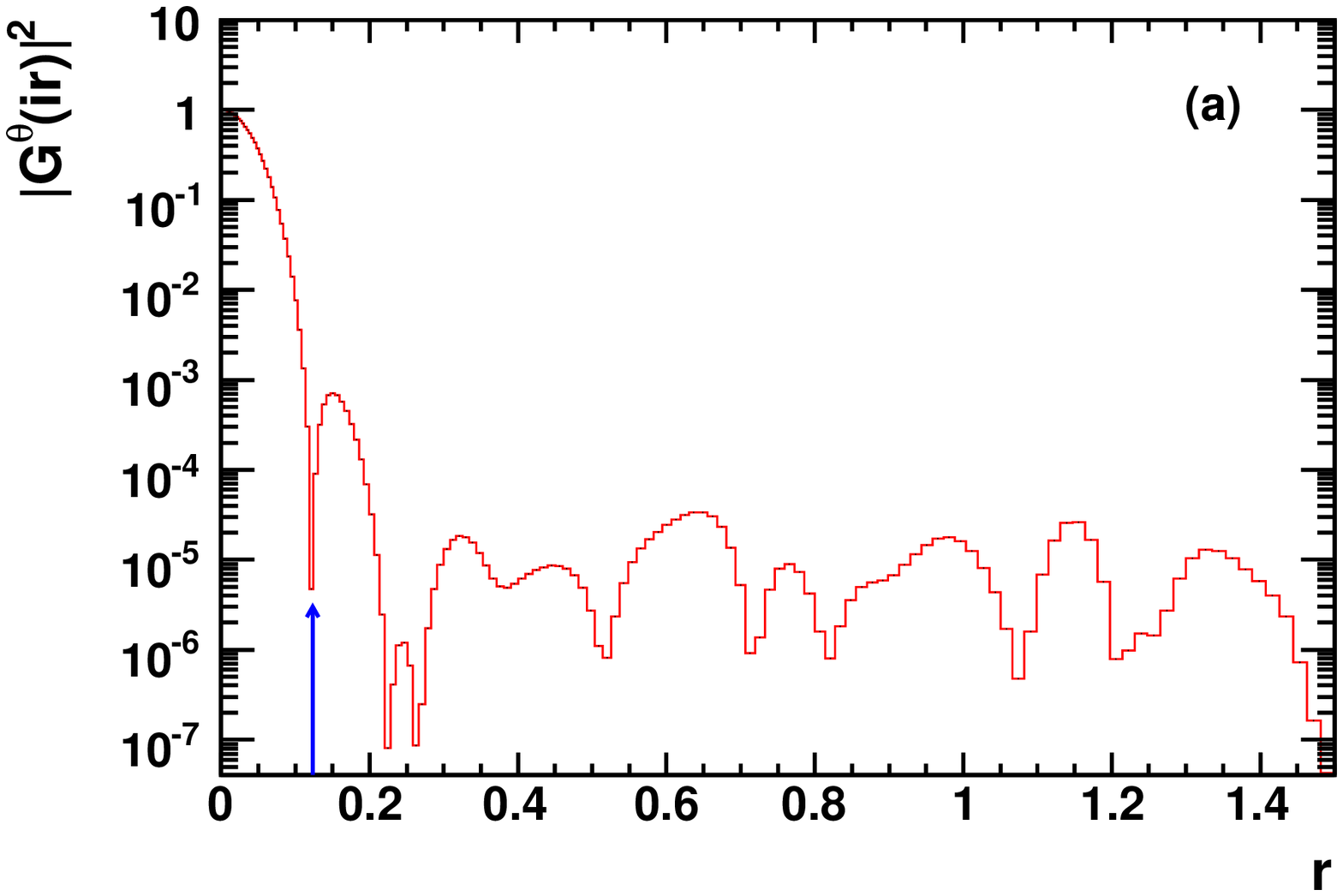}
\includegraphics[width=0.4\textwidth]{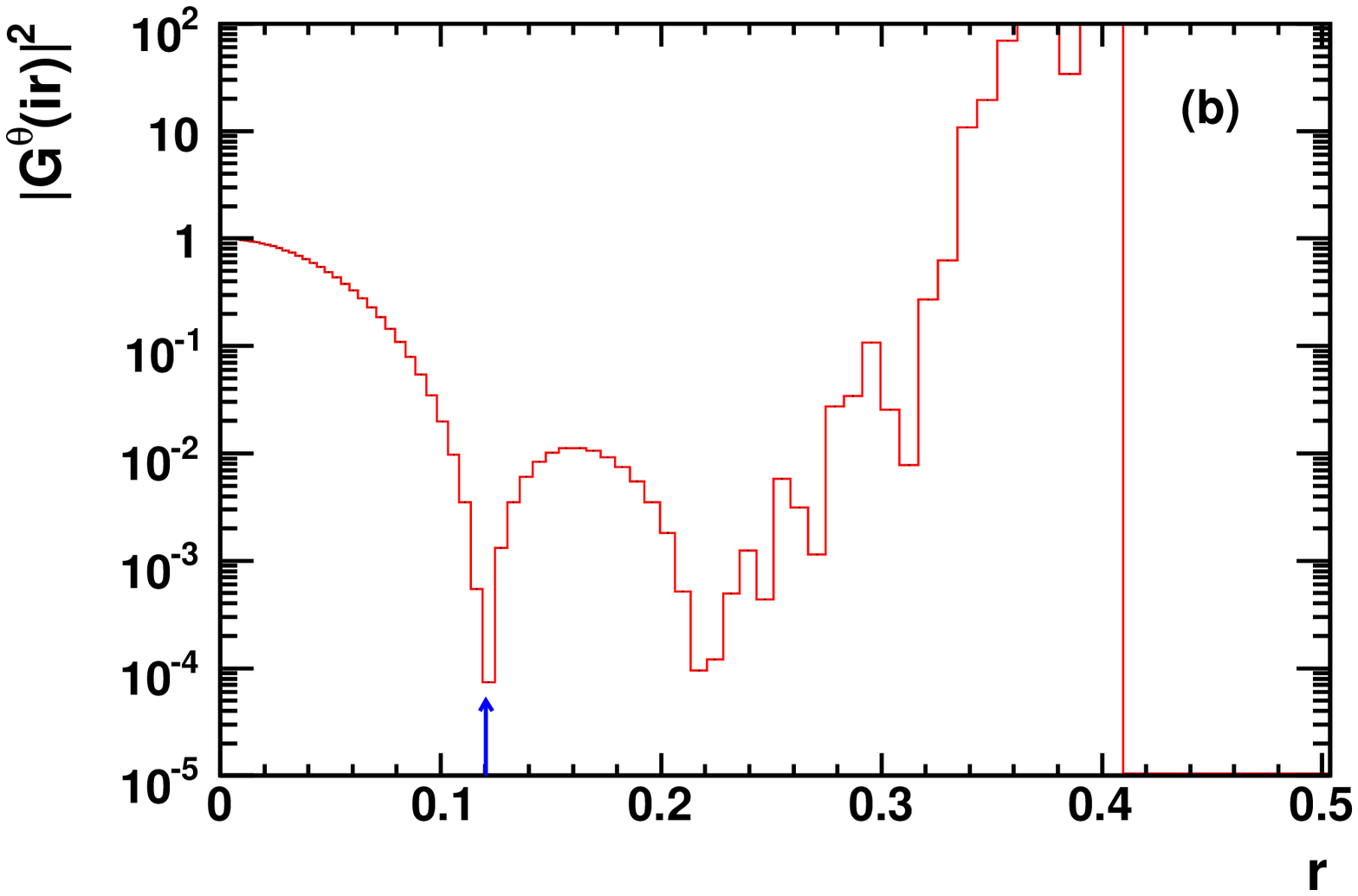}
\caption{(Color online) Examples of the modulus of the second
harmonic Lee-Yang Zero Generating Functions plotted as a function of
the imaginary axis coordinate, $r$.  The Sum Generating Function is
shown in (a) and the Product Generating Function in (b). The
vertical arrows indicate the positions of the first minimum, called
$r_0$. Note that in (b) the horizontal scale does not go out as far
because the calculations were terminated. All data are from \sqrtsNN
= 200 GeV \auau collisions.  }
\label{lyz}
\end{figure}

The Lee-Yang Zero method~\cite{LYZ_PLB,LYZ} is based on a 1952
proposal of Lee and Yang to detect a liquid-gas phase transition. As
opposed to the four-particle cumulant method which is sensitive
to the correlations of four particles, this method is sensitive to
the correlations of all the particles. Thus it is supposed to remove
non-flow correlations to all orders. It has so far been used only to
analyze one set of experimental data~\cite{LYZ_FOPI} and one set of
transport calculations~\cite{LYZ_UrQMD}. The method utilizes the
second-harmonic flow vector, $Q_2$, projected on to an arbitrary
laboratory angle, $\theta$:
\begin{equation}
\label{Qtheta}
Q^\theta_2 = \sum_{j=1}^M w_j \cos(2(\phi_j - \theta)),
\end{equation}
where the sum is taken over all the particles $j$ with lab angles
$\phi_j$ and weights $w_j$. For this method the weights are taken to
be the value of \pt in \GeVc\: for unidentified charged hadrons and
1.0 for identified particles. We have taken five equally spaced values
of $\theta$ to average out detector acceptance effects. The results
were not different when 20 values of $\theta$ were used. The theory of
the method~\cite{LYZ_PLB} is to find a zero of a complex generating
function, but in practice the first minimum of the modulus of the
generating function along the imaginary axis is used. The sum
generating function based on $Q_2^\theta$ is given by
\begin{equation}
\label{Gtheta}
G^\theta_2(ir) = \ \mid \langle e^{\mathit{i}rQ^\theta_2}
\rangle \mid,
\end{equation}
where $r$ is a variable along the imaginary axis of the complex plane
and the average is taken over all events. When data are analyzed in
small batches, the $G^\theta_2(ir)$ histograms are combined before
finding the first minimum. Such a histogram is shown in
Fig.~\ref{lyz}(a).  The square of the modulus is used to determine the
first minimum. The position along the imaginary axis of the first
minimum of the modulus of the generating function at the lab angle
$\theta$ is called $r_{0}^{\theta}$, and is related to the
``integrated'' flow by
\begin{equation}
\label{Vtheta}
V_2^\theta = \mathit{j}_{01} / r_0^\theta
\end{equation}
\begin{equation}
\label{v}
v_2 = \mean{V_2^\theta}_\theta / M,
\end{equation}
where $\mathit{j}_{01} = 2.405$ is the first root of the Bessel
Function $J_0$ and $M$ is the multiplicity. In the second equation the
average is taken over the lab angles $\theta$. However, Eq.~(\ref{v})
is only valid for unit weights. Normally, the anisotropic flow
parameter averaged over \pt and \et is obtained by taking the
yield-weighted average of the differential flow. For unit weights this
has been shown to agree with the ``integrated'' flow from
Eq.~(\ref{v}). The differential flow obtained by a second pass through
the data is given by
\begin{eqnarray}
\label{v_theta} v_{2m}^\theta(\eta,p_T) &=& V_2^\theta
\frac{J_1(j_{01})}{J_m(j_{01})}{\bf Re} \left(
\frac{\mean{\cos[2m(\phi_j - \theta)]e^{\mathit{i} r_0^\theta Q_2^\theta}}}
{\mathit{i}^{m-1} \mean{Q_2^\theta e^{\mathit{i} r_0^\theta Q_2^\theta}}} \right), \\
v_{2m}(\eta,p_T) &=& \mean{v_{2m}^\theta(\eta,p_T)}_\theta \nonumber
\end{eqnarray}
where $m=1$ for $v_2$ and $m=2$ for $v_4$. The average in the
numerator in the first equation is over the particles of interest
and the average in the denominator is over all events.

The product generating function is
\begin{equation}
\label{GthetaProd}
G^\theta_2(ir) = \ \mid \langle \prod_{j=1}^M [ 1 + \mathit{i}r
w_j \cos(2(\phi_j - \theta))] \rangle \mid .
\end{equation}
This takes more computer time because the product over all particles
has to be calculated for each value of $r$. An example is shown in
Fig.~\ref{lyz}(b). It can be seen that the sum generating function
oscillates after the first minimum, but the product generating
function rises very fast. Thus, for the product generating function,
the calculation was halted when $|G^\theta_2(ir)|^2$ got larger than
1000. This happened at various $r$ values between 0.2 and 0.4.  While
the method using the sum generating function is slightly faster than
the standard method~\cite{v2Methods}, using the product generating
function is about four times slower. For the product generating
function, Eqs.~(\ref{Vtheta}, \ref{v}) still hold, but the
differential flow is given by
\begin{equation}
\label{v_thetaProd}
v_{2m}^\theta(\eta,p_T) = V_2^\theta \frac{J_1(j_{01})}{J_m(j_{01})}{\bf Re}
\left( \frac{\left\langle G_2^\theta(\mathit{i} r_0^\theta) \frac{\cos(2m(\phi_j - \theta))}
{1 + \mathit{i}r_0^\theta w_j \cos(2(\phi_j - \theta))} \right\rangle
} { \mathit{i}^{m-1} \left\langle G_2^\theta(\mathit{i} r_0^\theta)
\sum_j \frac{w_j \cos(2(\phi_j - \theta))} {1 + \mathit{i}r_0^\theta w_j
\cos(2(\phi_j - \theta))} \right\rangle } \right) ,
\end{equation}
where again the average in the numerator is over the particles of
interest and the average in the denominator is over all
events. Although the sum generating function works fine for $v_2$,
analyses for $v_4$ (and $v_1$) have to be based on the product
generating function~\cite{LYZ}. This is because the product generating
function is better at suppressing autocorrelation effects which are
more important for mixed harmonics. All methods used in this paper
have been tested on simulated data. Also, since drift of the beam in
the detector over time might simulate the effect of anisotropic flow,
run-by-run recentering of the $Q$ vector was applied, but produced no
improvement in the results.

The errors were calculated from the variation of the results for
different event sub-samples. For very large errors this technique
could underestimate the error because even when there is no flow the
method will find a minimum from a fluctuation. In fact the Lee-Yang
Zero method only works for sufficient signal-to-noise ratio. Since the
signal is $v_2$ and the noise is proportional to $1/\sqrt{M}$, the
parameter $\chi = v_2 \sqrt{M}$ determines the applicability of the
method. We find that the errors get large and the results scatter when
$\chi \lt 0.8$, and thus the results are presented here only for
10--50\% centrality. The method fails for more central collisions
because $v_2$ is small, and for more peripheral collisions because the
multiplicity is small.

\section{Results}
\label{sect_result}

\subsection{Charged hadrons}

\begin{figure}[ht]
\includegraphics[width=0.5\textwidth]{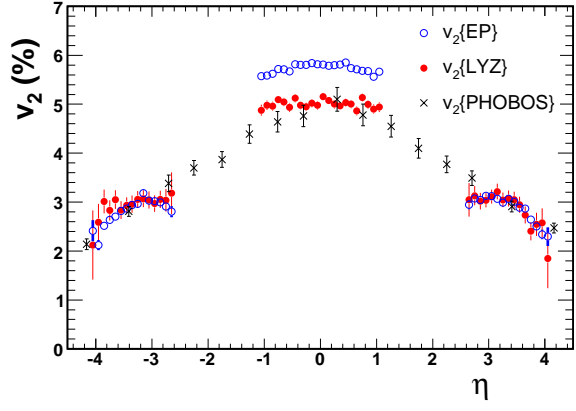}
\caption{(Color online) $v_2$ for charged hadrons from the Lee-Yang
Zero Product Generating Function (solid circles) and from the
Event-Plane method (open circles), as a function of pseudorapidity.
Both sets of data have been averaged over \pt from 0.15 to 2.0 \GeVc\:
and centrality from 10 to 40\% of \sqrtsNN = 200 GeV \auau
collisions.  For comparison, the PHOBOS data (10--40\%)~\cite{phobos}
(crosses) are also shown. The error bars are shown only for the
statistical uncertainties.} \label{v2LYZeta}
\end{figure}

\begin{figure}[ht]
\includegraphics[width=0.6\textwidth]{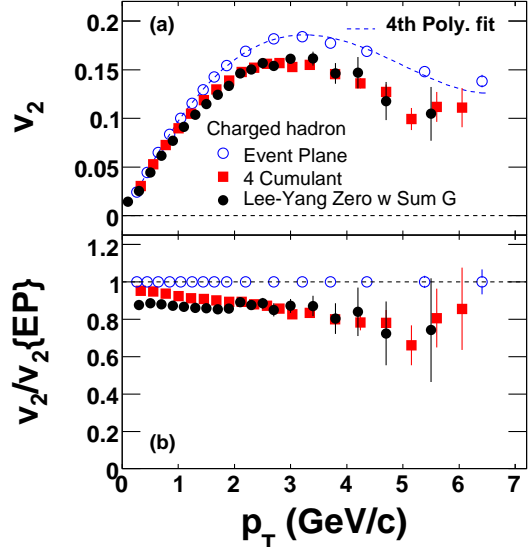}
\caption{(Color online) (a) $v_2$ as a function of \pt for charged
hadrons with $\mid \eta \mid \lt 1.0$ in 10--40\% \auau collisions, at
\sqrtsNN = 200 GeV, from the Event-Plane method (open circles),
4-particle cumulant method (solid squares), and Lee-Yang Zero method
(solid circles) with Sum Generating Function. (b) The ratios to the
polynomial fit to $v_2$\{EP\} are shown for $v_2$\{4\}/$v_2$\{EP\} and
$v_2$\{LYZ\}/$v_2$\{EP\} as a function of transverse momentum. The
error bars are shown only for the statistical uncertainties.}
\label{v2pt}
\end{figure}

\begin{figure}[ht]
\includegraphics[width=0.5\textwidth]{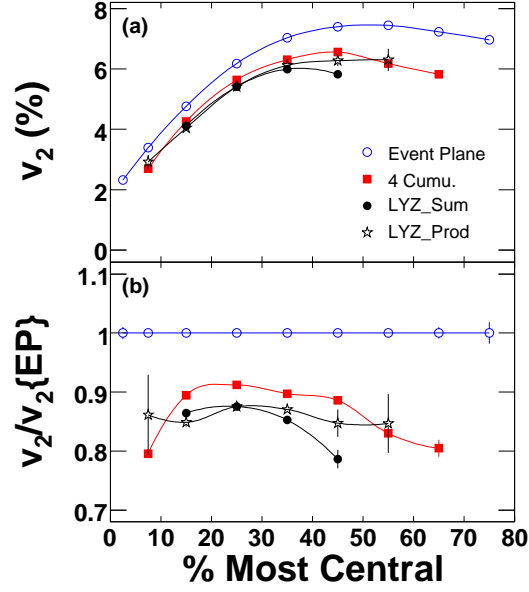}
\caption{(Color online) (a) $p_T$-integrated charged hadron $v_2$ in
the TPC as a function of geometrical cross section. Shown are the
Event-Plane method ($v_2$\{EP\}) (open circles), Lee-Yang Zero method
with Sum Generating Function (solid circles), Lee-Yang Zero method
with Product Generating Function (open stars), and 4-particle cumulant
method ($v_2\{4\}$) (solid squares). For the TPC, $\mid\eta\mid < 1.0$
was used, except for Lee-Yang Zero method with Product Generating
Function where the \et limit went to 1.3. (b) $v_2$ divided by
$v_2$\{EP\}. All data are from \sqrtsNN = 200 GeV \auau
collisions. The error bars are shown only for the statistical
uncertainties.} \label{intv2}
\end{figure}

To evaluate the different flow analysis methods and to estimate
systematic uncertainties, charged hadrons were analyzed first.
Figure~\ref{v2LYZeta} shows $v_2(\eta)$ for both the Lee-Yang Zero and
Event-Plane methods. For the Event-Plane method, the event plane was
taken from the main TPC, both for tracks in the TPC as well as the
FTPC. For the Lee-Yang Zero method, where there is no event plane,
tracks in all three TPCs were used. The Lee-Yang Zero results are for
the product generating function but are in agreement with the sum
generating function results. Elliptic flow falls off in the Forward
TPC (FTPC) covering $\mid \eta \mid$ from 2.6 to 4.2. This fall off is
probably because the spectra as a function of \pt are steeper at high
\et and give less weight to the large $v_2$ values at high
$p_T$~\cite{BRAMSeta}. Agreement in the FTPC region between
two-particle and multiparticle methods has been seen
previously~\cite{flowPRC}. Having a gap in pseudorapidity between the
particles being correlated reduces the non-flow effects due to short
range correlations. Indeed, PHOBOS correlates particles with an event
plane from a different part of their detector, which is essentially
similar to the $\eta$-subevent method. With $\mid \eta \mid \le 1$,
PHOBOS~\cite{phobos} data points are consistent with STAR Lee-Yang
Zero data although it appears that PHOBOS data may be more
peaked. Averaging over the TPC \et region $\mid \eta \mid \lt 1.0$,
the $v_2(p_T)$ values are shown in Fig.~\ref{v2pt} (a) together with
Event-Plane and 4-particle cumulant results.  For these charged
hadrons, the ratio of the four-particle cumulant result to the
Event-Plane method shown in Fig.~\ref{v2pt} (b) falls off as \pt
increases.  This indicates a non-flow effect in the Event-Plane method
which increases with \pt as one would expect for the contribution of
jets. On the other hand, the Lee-Yang Zero ratio seems to be
flat. Figure~\ref{intv2} (a) shows the integrated $v_2$ (averaged over
\et and $p_T$) for four different analysis methods as a function of
centrality. The ratios to the Event-Plane method are shown in
Fig.~\ref{intv2} (b). The centrality region where all methods have
reasonable error bars is 10 to 50\%. The Event-Plane method appears to
be about 15\% higher compared to the other methods known to greatly
reduce non-flow effects. This effect was already seen in the
differential data as a function of $\eta$ for $p_T \lt 2$ \GeVc\: in
Fig.~\ref{v2LYZeta} and as a function of $p_T$ in Fig.~\ref{v2pt}. For
the most peripheral collisions non-flow might be larger and for the
most central collisions fluctuations could be important. From
Fig.~\ref{intv2} we would estimate the systematic errors at these
other centralities to be 20\%, and also probably this same value for
minimum bias events.

\clearpage
\subsection{\ks and \lam}

\begin{figure}[ht]
\includegraphics[width=0.65\textwidth]{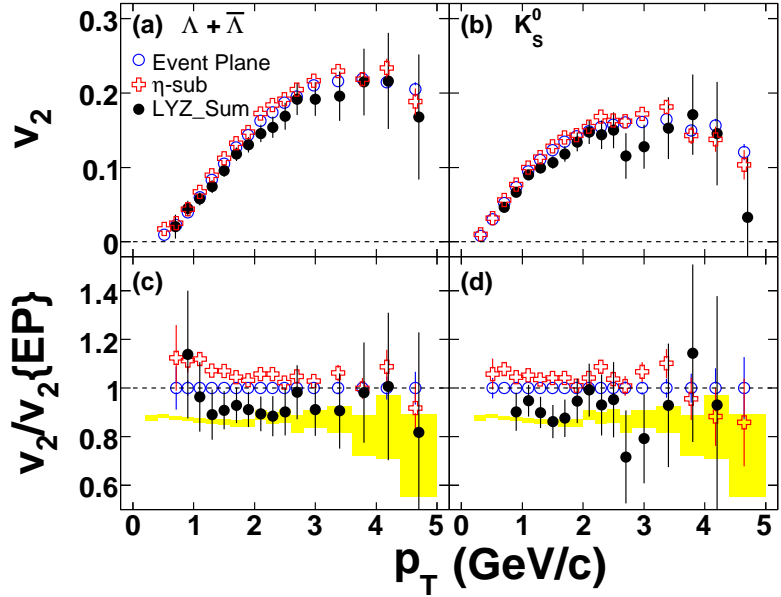}
\caption{(Color online) $v_2$ as a function of \pt for 10--40\%
centrality using Event-Plane method (open circles), Lee-Yang Zero
method with Sum Generating Function (solid circles), and
$\eta$-subevent method (open crosses), are shown in (a) and (b) for
$\Lambda$ and $K_{S}^{0}$, respectively. All data are from \sqrtsNN
= 200 GeV \auau collisions. The ratios, $v_2$\{LYZ\}$/v_2$\{EP\} and
$v_2\{\eta\}/v_2$\{EP\}, are shown in (c) and (d) for $\Lambda$ and
$K_{S}^{0}$, respectively. $v_2$\{LYZ\}$/v_2$\{EP\} ratios for charged
hadrons are shown as shaded bands. The error bars are shown only for
the statistical uncertainties.} \label{kslav2}
\end{figure}

In order to estimate the particle dependence of systematic errors,
\ks mesons and \lam baryons were analyzed with different flow
analysis methods.  Figure~\ref{kslav2} shows 10--40\% $v_2(p_T)$ of (a)
\lam + \alam and (b) \ks obtained with the Event-Plane, Lee-Yang
Zero, and $\eta$-subevent methods. Ratios of $v_2$ from these various
methods to the Event-Plane method are shown in Fig.~\ref{kslav2} (c)
and (d) for \lam + \alam and \ks, respectively. The results from
Lee-Yang Zero method are about 10\% lower than those from the
Event-Plane method. This is consistent with charged particles,
indicating that non-flow effects for \ks and \lam are also reduced in
the Lee-Yang Zero method. However, the ratio of the Lee-Yang Zero to
Event-Plane method appears to be flat up to 5 \GeVc, considering the
large statistical uncertainties. This is again similar to the trend
observed for charged particle Lee-Yang Zero results, which are shown
as shaded bands. The results for the $\eta$-subevent method are higher
than for the Event-Plane method especially at low $p_T$. The
$v_2\{\eta\}/v_2$\{EP\} ratio can be better understood by factorizing
the ratio into an observed $v_2$ term and a resolution term
$(v_{2}^{obs}\{\eta\}/v_{2}^{obs}$
\{EP\})$\times(resolution$\{EP\}$/resolution\{\eta\})$. In this
analysis we calculate the $\eta$ subevent resolution by correlating
the event planes from the different $\eta$ hemispheres. In this
case, the non-flow effects are reduced in both $v_2^{obs}$ and the
resolution. These factors contribute in the opposite direction to
$v_{2}\{\eta\}/v_{2}$\{EP\}; non-flow in the resolution term
increases the ratio while non-flow in the $v_{2}^{obs}$ term
decreases the ratio. The $v_{2}\{\eta\}/v_{2}$\{EP\} ratio is
greater than unity because the resolution is more sensitive to
non-flow than $v_{2}^{obs}$ and the decrease with $p_T$ is caused by
the increase of non-flow effects in $v_{2}^{obs}$\{EP\} with
increasing $p_T$.

\subsection{Systematic uncertainty of $\Lambda$ from feed-down}

We estimate the reaction plane orientation from the azimuthal
distribution of charged particles measured with the TPC ($|\eta|<1.3$),
constructing the second harmonic event plane flow vector $Q_2$.  The
TPC is also used to reconstruct $\Lambda$ and $\overline\Lambda$ hyperons
via their charged decay daughters, $\pi^{\pm}$ or $p (\overline p)$.
Elliptic flow of $\Lambda$ and $\overline\Lambda$ is measured by
correlating the hyperon azimuthal angle with the event plane vector
$Q_2$.  Although the correlation strength is mainly defined by the
hyperon elliptic flow, $v_2^{\Lambda,\overline\Lambda}$, in such an
approach other sources of correlations (non-flow effects) may
contribute and bias the measured $v_2^{\Lambda,\overline\Lambda}$ values.

We do not distinguish between $\Lambda$ and $\overline\Lambda$
particles produced from the secondary decays (for example,
$\Xi^{-}\to\Lambda+\pi^{-}$
($\overline\Xi^{+}\to\overline\Lambda+\pi^{+}$),
$\Sigma(1385)^{-}\to\Lambda +\pi^{-}$
($\Sigma(1385)^{+}\to\overline\Lambda+\pi^{+}$) or
$\Sigma^{0}\to\Lambda +\gamma$ ) and hyperons which originate directly
from the primary interaction. Indirect hyperons lead to the presence
of extra correlations that are not related to the reaction plane
between hyperons and other charged particles produced in the
collision. Note that the charge combinations for these correlations
are opposite for $\Lambda$ and $\overline\Lambda$ particles.

To estimate the contribution of these non-flow correlations from
hyperon feed-down effects we use the charge subevent technique. For
this method we introduce two event plane vectors: $Q^{+}_2$
constructed from positively charged particles and $Q^{-}_{2}$ from
negatively charged particles. We then estimate the contribution to the
$\Lambda$ and $\overline\Lambda$ elliptic flow measured with the full
event plane vector $Q_{2}$ by considering the following ratio:

\begin{eqnarray}
\label{non-flowFromFeedDown}
\delta R\{\rm FeedDown\} = \frac{v_2^{\Lambda}\{Q^{-}_{2}\}
+ v_2^{\overline\Lambda}\{Q^{+}_{2}\}} {v_2^{\Lambda}\{Q^{+}_{2}\}
+ v_2^{\overline\Lambda}\{Q^{-}_{2}\}}.
\end{eqnarray}

Here $v_2^{\Lambda,\overline\Lambda}\{Q^{\pm}_{2}\}$ denotes $\Lambda$
($\overline\Lambda$) elliptic flow values measured from correlations with
$Q^{\pm}_{2}$.  The numerator in Eq.~(\ref{non-flowFromFeedDown})
contains the contributions from non-flow correlations attributed to
feed-down effects, while the denominator is free of this.  From this
study we found that contribution of non-flow effects from feed-down of
secondary $\Lambda$ and $\overline\Lambda$ hyperons is $\leq 2\%$.

\subsection{Other systematic uncertainties}

In order to estimate the systematic uncertainties in the identified
hadron $v_2$, we employed the standard Event-Plane method, the
$\eta$-subevent method, and the Lee-Yang Zero method. The results of
the analysis for $\Lambda$+\alam and \ks are shown in
Fig.~\ref{kslav2}.  The ratios to the event plane results are shown in
the lower panels. We limited ourselves to the 10--40\% centrality bin
where the results from the Lee-Yang Zero method are most reliable. For
comparison, the charged hadron results from the
$v_2$\{LYZ\}$/v_2$\{EP\} ratios (Fig.~\ref{v2pt}) are also shown in
the figure as shaded bands.  There is no clear $p_T$ dependence of the
ratio $v_2$\{LYZ\}$/v_2$\{EP\} for either $\Lambda$+\alam in (c) or
for \ks in (d) although, within the statistical errors, similar trends
are seen for both identified particles. The overall systematic errors
are on the order of 15\% in the $p_T$ region studied. The subevent
method, however, introduced an additional factor described in
Section~\ref{etaSubs} that leads to the enhanced ratio for both
$\Lambda$+\alam and $K_S^0$. This opposite effect is also within the
order of 15\%.

In order to obtain good statistics, the Event-Plane method is used for
most of analyses of identified particles. Depending on the analysis
method, systematic uncertainties from variations in particle
identification cuts, background subtractions, and summing of
centrality bins are also estimated. By varying particle identification
(PID) cuts, which change signal over background ratios by a factor of
3, the systematic uncertainty from the PID cuts is estimated to be
about 5\% below 4 \GeVc. From 4 \GeVc\: to 6 \GeVc, this effect is
larger in central collisions than peripheral collisions, and in the
0--10\% bin it is about a 10\% effect. To estimate the uncertainty
from background subtractions, background variations from different
second and fourth order polynomial fit functions in Fig.~\ref{v2vsm}
were propagated to measured $v_{2}$ values with Eq.~\ref{v2minv}. The
effect is less than 3\%. When combining centralities, the combined
$v_{2}$ value should be a yield-weighted average of $v_2$ values in
small centrality bins. In the method used previously~\cite{v2Methods},
$v_2$ values are taken as weighted observed $v_2$ corrected by
weighted event plane resolution. With the $v_2$ versus $m_{inv}$
method, similar corrections were calculated. They are less than 5\%
below 6 \GeVc.

In summary, for charged particles, a 15\% difference for $\mean{v_2}$
at mid-rapidity for 10--40\% collisions between the event plane and
the Lee-Yang Zero methods has been observed. The difference between
$v_2$\{EP\} and $v_2\{4\}$ is smaller, $\sim$10\%, but for the more
peripheral and more central collisions it seems to be closer to
20\%. For $v_2(p_t)$ of \ks and \lam\!, a difference also is observed
between the Event-Plane and the Lee-Yang Zero methods. However, the
comparison with charged particles shows that, within the much larger
statistical uncertainties for the \lam and \ks analysis, the different
magnitudes of the estimated flow from the different methods observed
are similar for charged particles and for the \lam and \ks
particles. The uncertainty used is 15\%. In a latter section, the
results are presented for \xim and the \omm\!. For these particles,
due to limited statistics, only the Event-Plane method has been used
and therefore no real estimate of the systematic error is
available. Instead what was done, was to show the estimated systematic
uncertainties obtained for the charged particles, taking into account
effects from background estimation, summing centralities, and
variations in cuts. A summary of our best knowledge of systematic
errors is given in Table~\ref{tab:errors}.

\begin{table}[ht]
\centering
\begin{tabular}{|c||c|c|c|c|} \hline
     centrality           & 0--80\%   & 40--80\% & 10--40\% & 0--10\% \\ \hline \hline
     charged hadrons      & 20\%      & 20\%     & 15\%     & 20\%  \\ \hline
     identified particles & N/A     & N/A      & 15\%     & N/A   \\ \hline
\end{tabular}
\caption{Systematic errors of $v_2$ estimated from the different flow
methods summarized as a function of centrality for $\sqrtsNN = 200$
GeV \auau collisions for unidentified charged hadrons and identified
particles. The \pt region covered for charged hadrons is 0.5 $\lt$ \pt
$\lt$ 7.0 \GeVc\: and for identified particles $K_S^0$, $p$, and
$\Lambda$ is 0.5 $\lt$ \pt $\lt$ 5.0 \GeVc.}
\label{tab:errors}
\end{table}

\section{Discussion}
\label{discus}

\subsection{Charged hadrons}

Figure~\ref{intv2} shows that for the Lee-Yang Zero method the sum and
product generating functions agree, but are slightly lower than the
four-particle cummulant method. The Event-Plane method appears to be
about 15\% higher than the other methods. This could be due to either
non-flow contributions increasing the event plane results or
fluctuations of $v_2$ decreasing the multi-particle
methods~\cite{flowPRC}. Charged hadron results give an indication of
the systematic uncertainty inherent in the Event-Plane method, which
is used for most of the identified particles in this paper in order to
reduce statistical errors. The \pt dependence of this effect can be
seen in Fig.~\ref{v2pt}. Note that the mentioned 15\% non-flow effect
is extracted only from the 10--40\% centrality window. In other
centrality bins, the effect may be larger. It should be possible to
study this effect as a function of \pt for all centralities with a
future higher statistics data sample.

\subsection{Identified hadrons}

\begin{figure}[ht]
\center{\includegraphics[width=0.75\textwidth]{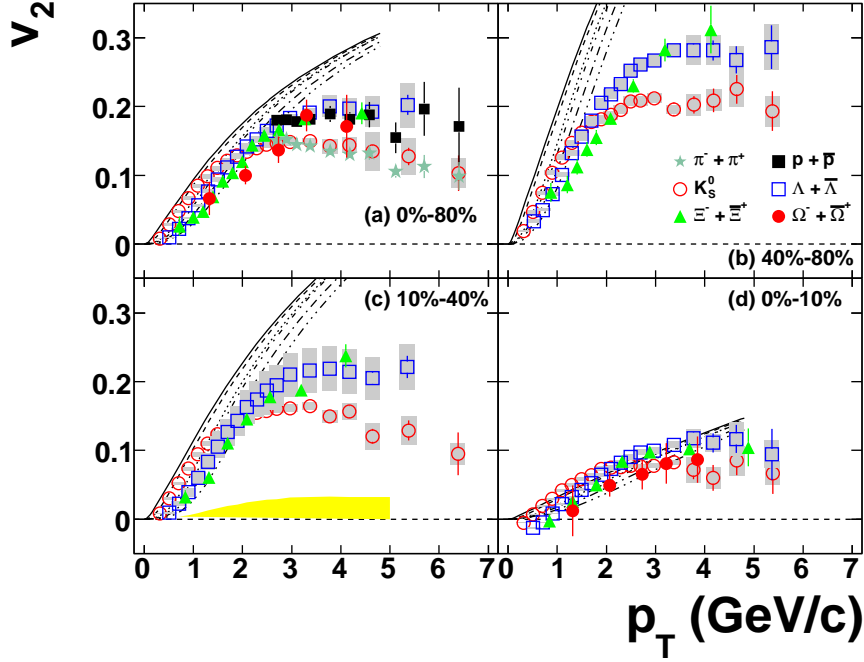}}
\caption{(Color online) $v_2$ of $K_S^0$ (open circles), $\Lambda$
(open squares), $\Xi$ (filled triangles) and $\Omega$ (filled circles)
as a function of \pt for (a) 0--80\%, (b) 40--80\%, (c) 10--40\% and
(d) 0--10\% in \auau collisions at \sqrtsNN = 200 GeV.  The error bars
represent statistical uncertainties. The bands on the data points
represent systematic uncertainties as discussed in the text. For
comparison, pion (stars) and proton (filled squares) results are shown
in (a). The systematic uncertainty of non-flow for $K_S^0$ and
$\Lambda$ for 10--40\% (c) is plotted as a shaded band near 0.  For
comparison, results from ideal hydrodynamic
calculations~\cite{pasi03,pasirev06} are shown: at a given $p_T$, from
top to bottom, the lines represent the results for $\pi$, $K$, $p$,
$\Lambda$, $\Xi$, and $\Omega$.} \label{mainv2pt}
\end{figure}

The results for $\pi^+$+$\pi^-$, $\overline{p}+p$, $K_S^0$,
$\Lambda$+\alam, $\Xi^-$+$\overline{\Xi}^+$ and
$\Omega^-$+$\overline{\Omega}^+$ are shown in Figure~\ref{mainv2pt}
for various centralities of \auau collisions at 200 GeV. Shown are
results for minimum bias and three other centrality bins. All
$v_2(p_T)$ results are from the Event-Plane method. The systematic
uncertainties extracted from PID cuts, background subtractions, and
combining centralities, are shown as shaded bars in the figure. The
systematic uncertainty in the method itself is not included. The
shaded band in plot (c) indicates the systematic uncertainties for
\ks~and $\Lambda$ for the 10--40\% centrality bin, as discussed in the
previous section. In addition, the systematic uncertainties for pions
and protons from Event-Plane and $\eta$-subevent analysis are plotted
as the shaded band in plot (a).  The results from an ideal
hydrodynamic model~\cite{pasi03,pasirev06} are displayed by the lines.

Figure~\ref{mainv2pt} shows that the ideal hydrodynamic model
calculations reproduce the mass ordering of $v_2$ in the relatively
low $p_T$ region (the heavier the mass, the smaller the $v_2$) but
overshoot the values of $v_2$ for all centrality bins. There seems to
be a $p_T$ dependence in the disagreement, and for more central
collisions, the overshoot does not take place until a higher $p_T$. In
other words, the system agrees better with the ideal hydrodynamic
model for more central collisions. Although we do not expect a large
non-flow contribution at the low transverse momentum region, the
centrality selections between the model calculations based on the
impact parameter and the data based on the multiplicity are different,
which may also affect the model and data agreement.  Note that we
observe possible negative values of $v_2(p_T)$ for the heavier hadrons
at the lowest observed $p_T$ in the most central \auau collisions.

At higher $p_T$, the hydrodynamic type mass ordering evolves into a
hadron type ordering (baryons versus mesons). There the results show
two groups depending on the number of quarks in the hadron; the
baryons are higher than the mesons.  The effects are clearly shown in
plots (a), (b) and (c). In the most central bin (d), however, the
effect is less pronounced. For all $p_T$, $v_2$ evolves toward larger
values in going from central collisions to more peripheral
collisions. The ideal hydrodynamic model also predicts this centrality
dependence though it fails to describe the behavior at higher $p_T$.

\begin{figure}[ht]
\center{\includegraphics[width=0.65\textwidth]{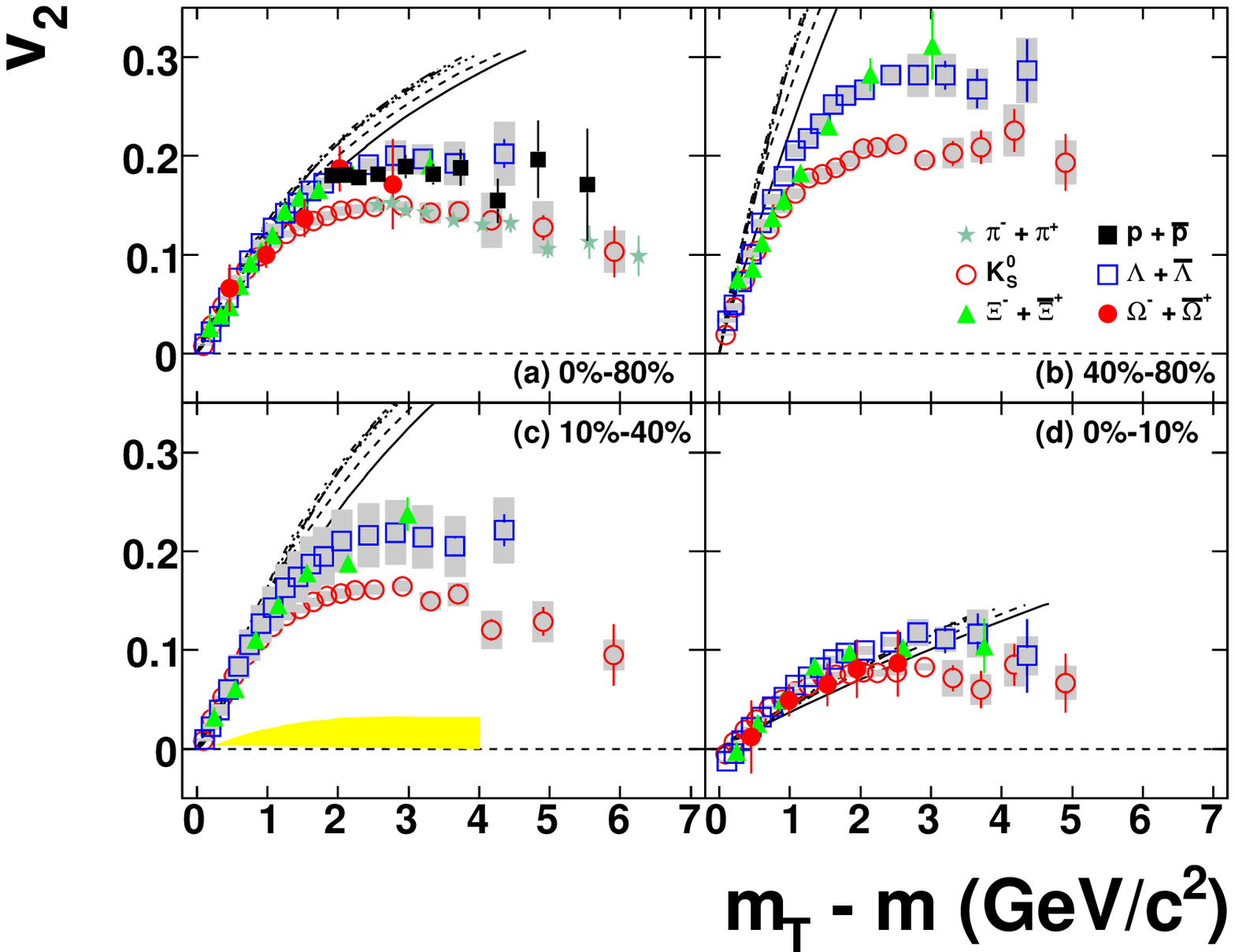}}
\caption{(Color online) $v_2$ in Fig.~\ref{mainv2pt} is re-plotted
as a function of $m_T-m$ for $K_S^0$ (open circles), $\Lambda$ (open
squares), $\Xi$ (filled triangles) and $\Omega$ (filled circles), for
(a) 0--80\%, (b) 40--80\%, (c) 10--40\% and (d) 0--10\% in \auau
collisions at \sqrtsNN = 200 GeV. For comparison, pion (stars) and
proton (filled squares) results are shown in (a). The ideal
hydrodynamic calculations in Fig.~\ref{mainv2pt} are also shown as a
function of $m_T-m$. } \label{mainv2mt}
\end{figure}

Figure~\ref{mainv2mt} shows the same results as in Fig.~\ref{mainv2pt}
but as a function of the transverse kinetic energy $K_T = m_T - m$ =
$\sqrt{p_T^2+m^2}-m$. Here {\it m} is the particle mass. In this case,
$v_2$ for all hadrons at low $K_T$ follow a universal curve, which
appears to be monotonically increasing and almost linear in all
centrality bins. The observed increase is slowest for the most central
0--10\% bin.  The corresponding results from the ideal hydrodynamic
model calculations are also shown in the figure. The mass ordering in
the model calculation is reversed when one plots $v_2$ versus $m_T-m$:
the higher the mass the larger the value of $v_2$. While the data seem
to show a scaling in the low $m_T-m$ region, the model results do not
show any scaling.

\begin{figure}[ht]
\center{\includegraphics[width=0.65\textwidth]{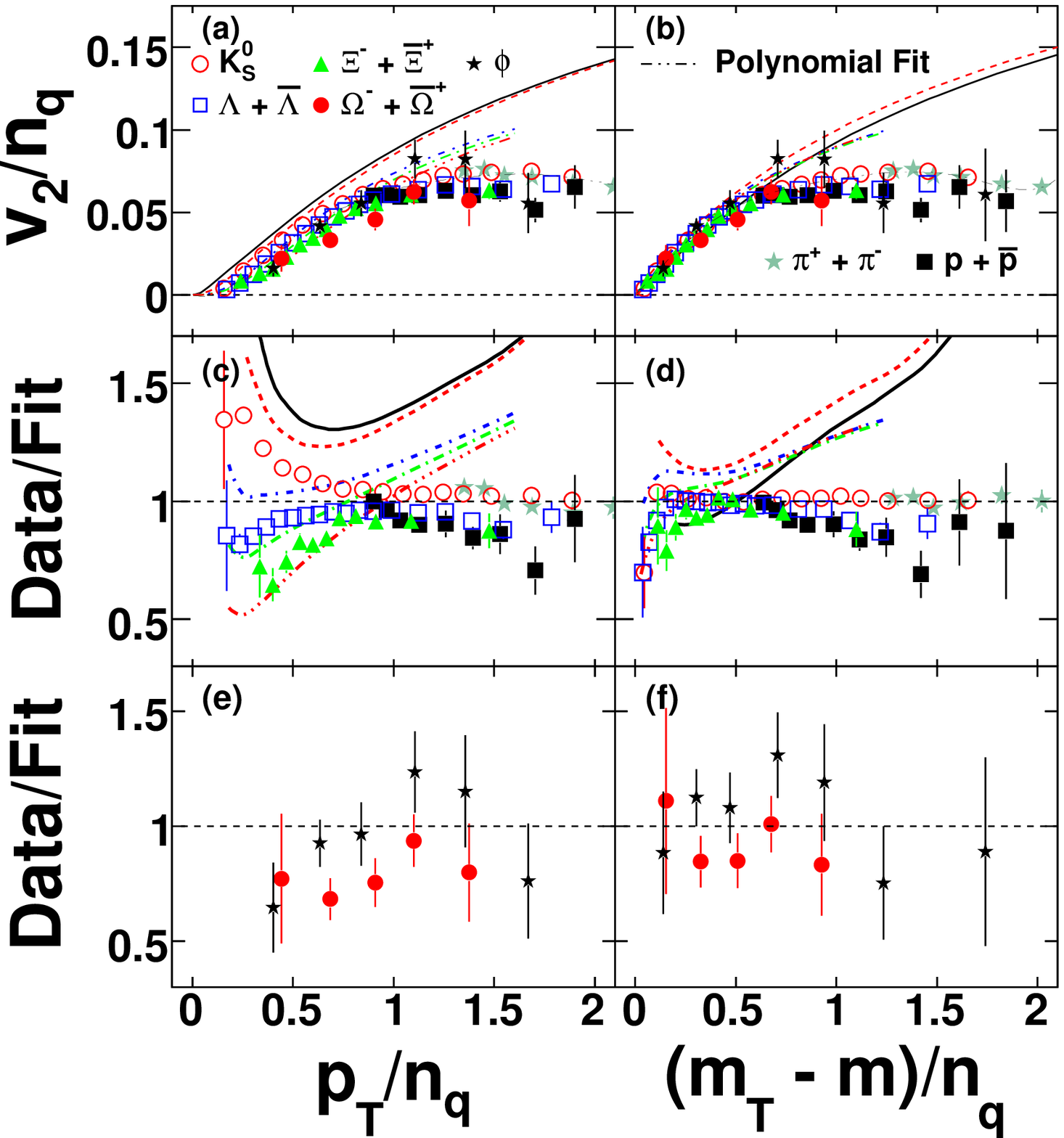}}
\caption{(Color online) Number-of-quark scaled $v_2$ ($v_2/n_q$) of
identified particles versus $p_T/n_q$ in $(\GeVc)$ (left column) and
($m_T-m$)/$n_q$ in $(\GeVc^2)$ (right column). Dot-dot-dashed lines
are the results of 6th order polynomial fits to $K_S^0$, $\Lambda$,
$\Xi$, and $\Omega$. The ratios of the data points over the fit are
shown in panels (c) and (d) for $K_S^0$, $\pi$, $p$, $\Lambda$, and
$\Xi$, and in panels (e) and (f) for $\Omega$ and
$\phi$~\cite{star_fv2}. The error bars are shown only for the
statistical uncertainties.  Ideal hydrodynamic calculations for $\pi$,
$K$, $\Lambda$, $\Xi$, and $\Omega$ are presented by solid lines,
dashed lines, dot-dashed lines, dot-long-dashed lines, and
dot-dot-dot-dashed lines, respectively.  The ratio of hydrodynamic
calculations over the fit are also shown for comparison in panels (c)
and (d). The data are from minimum bias (0--80\%) \auau collisions at
$\sqrtsNN = 200$ GeV.} \label{ncqv2}
\end{figure}

\begin{figure}[ht]
\center{\includegraphics[width=0.65\textwidth]{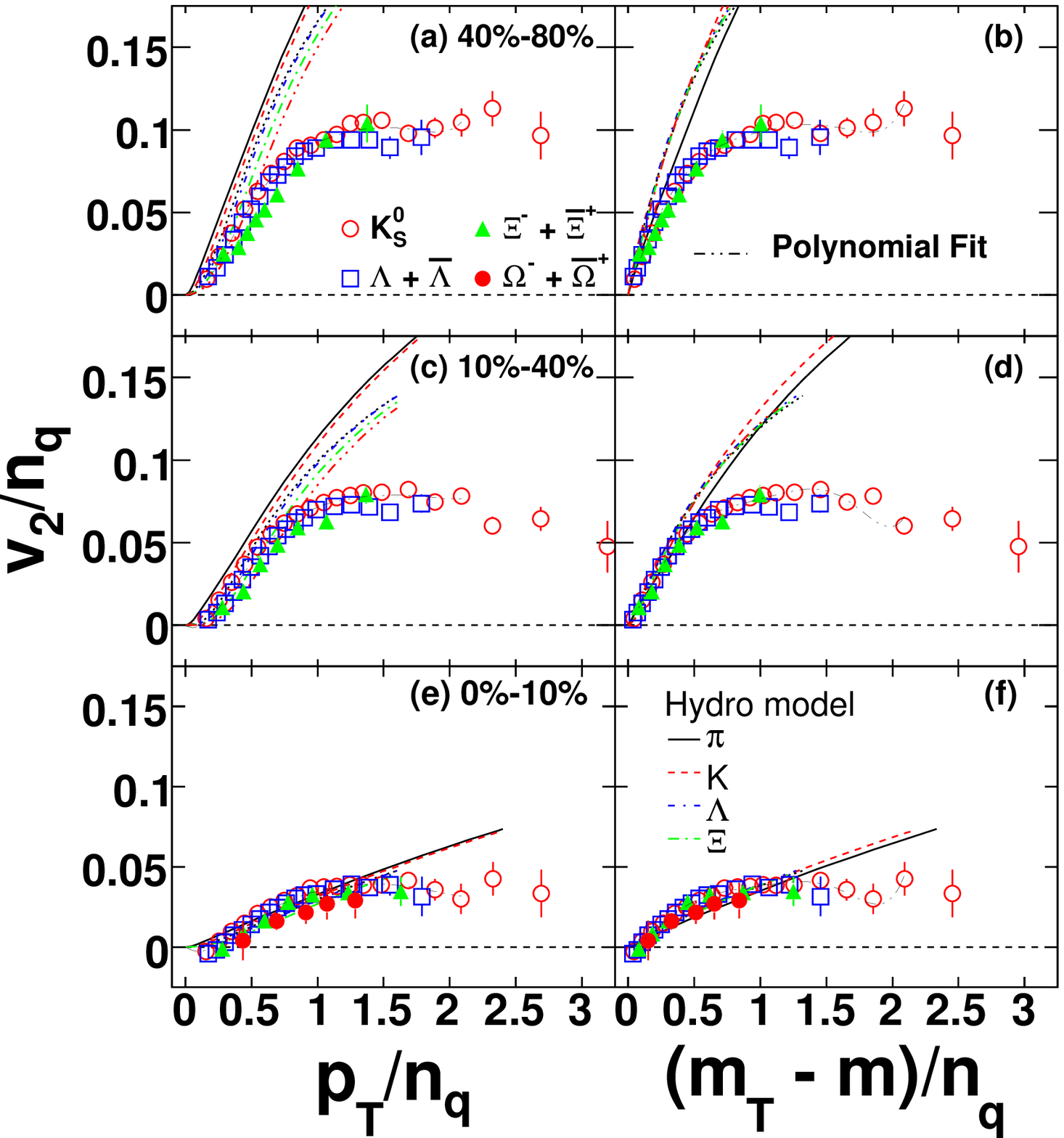}}
\caption{(Color online) Centrality dependence of the number-of-quark
scaled $v_2$ ($v_2/n_q$) of identified particles versus $p_T/n_q$ in
$(\GeVc)$ (left column) and ($m_T-m$)/$n_q$ in $(\GeVc^2)$ (right
column). The error bars are shown only for the statistical
uncertainties. The 6th order polynomial fits are shown as
dot-dot-dashed-lines. Ideal hydrodynamic curves~\cite{pasirev06} are
also plotted. All data are from \sqrtsNN = 200 GeV \auau collisions.}
\label{ncqv2cen}
\end{figure}

\begin{figure}[ht]
\center{\includegraphics[width=0.65\textwidth]{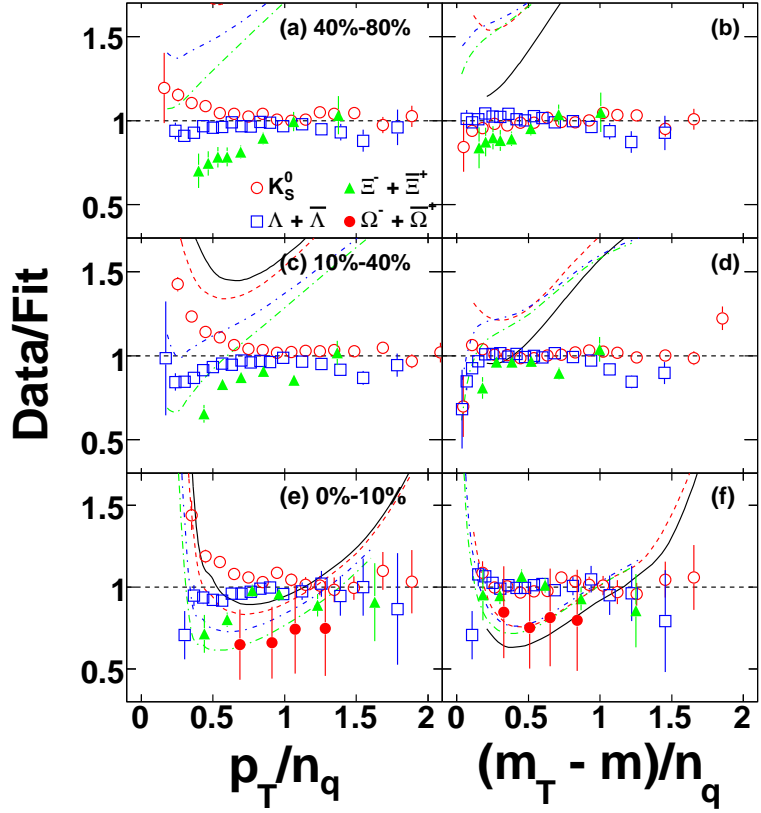}}
\caption{(Color online) Centrality dependence of the $v_2/n_q$ ratio
to a common polynomial fit is shown versus $p_T/n_q$ in $(\GeVc)$
(left column) and ($m_T-m$)/$n_q$ in $(\GeVc^2)$ (right column). The
error bars are shown only for the statistical uncertainties. Ideal
hydrodynamic calculations over the same fit are also shown for
comparison.  Ideal hydrodynamic calculations for $\pi$, $K$,
$\Lambda$, $\Xi$ and $\Omega$ are presented by solid lines, dashed
lines, dot-dashed lines, dot-long-dashed lines, and dot-dot-dot-dashed
lines, respectively.  All data are from \sqrtsNN = 200 GeV \auau
collisions.}
\label{ncqv2cenratio}
\end{figure}

In Figs.~\ref{ncqv2}, \ref{ncqv2cen}, and \ref{ncqv2cenratio} we
discuss the properties of the centrality dependence of the observed
scaling including both the low $p_T \le 2$ \GeVc\: and the
intermediate $2 \le p_T \le 5$ \GeVc\: regions. Figure~\ref{ncqv2}
shows $v_2$ scaled by the number of constituent quarks, $v_2/n_q$, for
all strange hadrons including the pure multi-strange hadrons $\phi$
and $\Omega$. The left panels show the results as a function of \pt
scaled by the number of quarks, $p_T/n_q$, and the right panels as a
function of $K_T/n_q$.  Plots (a) and (b) are the corresponding scaled
results. It appears that the scaling works better when the data are
plotted as a function of transverse kinetic energy $K_T$, as in plot
(b). The ideal hydrodynamic results are also shown in both
presentations. Clearly, the hydrodynamic distributions are also better
scaled when plotted versus $K_T$. Polynomial fits were made for all
hadrons. The results are shown as dot-dot-dashed lines in plot (a) and
(b). The ratios of the data and the hydrodynamic lines over the
polynomial fit are shown in plots (c) and (d). For all data, there is
scaling at $p_T/n_q \ge 0.7$ \GeVc\: or $K_T/n_q \ge 0.2$
\GeVc$^2$. The errors from the multi-strange hadrons
$\phi$~\cite{star_fv2} and $\Omega$ are large, see plots (e) and (f),
but are consistent with the scaling.  This observed $n_q$-scaling
provides strong evidence that these hadrons are formed via a
coalescence process at the end of the partonic
evolution~\cite{starwp,mv03}. Comparing to the light non-strange
hadrons, strange hadrons participate much less in later stage hadronic
rescattering processes~\cite{hsx}; thus, these distributions directly
reflect the early dynamics of the collision at RHIC. It is interesting
to note that the ideal hydrodynamic results scale neither at low $p_T$
nor intermediate $p_T$. Therefore the observed scaling $can not$ be a
general characteristics of hydrodynamic model
calculations~\cite{nagle}, although such calculations do show the
observed mass ordering in the low $p_T$ region.

Figures~\ref{ncqv2cen} and \ref{ncqv2cenratio} show the centrality
dependence of the scaling properties and the ratios, respectively.
Similar to the observations from Fig.~\ref{ncqv2}, the conclusions
from Figs.~\ref{ncqv2cen} and \ref{ncqv2cenratio} are:
\begin{enumerate}

\item There is a clear number-of-quark scaling at intermediate $p_T$
and better scaling in $K_T$ for all hadrons studied here, but no
scaling is observed at low $p_T$.

\item The ideal hydrodynamic model results do not show any scaling over the
region $0.2 \le p_T \le 5$ \GeVc.

\item These results are true for all centrality bins.

\end{enumerate}

\clearpage
\begin{table}[ht]
\centering
\begin{tabular}{|c||c|c|c|c|c|} \hline
    & $\sigma_{trig}/\sigma_{geom}$ & 0--80\%     & 40--80\%     & 10--40\%     & 0--10\% \\ \hline \hline
\multirow{5}{16mm}{62.4 GeV}
    & $\varepsilon_{part}$    & 0.3919$\pm$0.0003 & 0.5426$\pm$0.0004 & 0.2927$\pm$0.0003  & 0.1108 $\pm$0.0002 \\
    & $N_{part}$              & 122$\pm$3     & 39$\pm$5      & 167$\pm$7      & 320$\pm$3   \\
    & $N_{bin}$               & 249$\pm$13    & 50$\pm$10     & 338$\pm$18     & 797$\pm$9   \\ \hline
\multirow{5}{16mm}{200 GeV}
    & $\varepsilon_{part}$  & 0.3843$\pm$0.0001 & 0.5343$\pm$0.0002 & 0.2829$\pm$0.0001 & 0.1054$\pm$0.0001 \\
    & $N_{part}$                    & 126$\pm$8  & 42$\pm$7    & 173$\pm$10  & 326 $\pm$ 6  \\
    & $N_{bin}$                     & 293$\pm$36 & 57 $\pm$ 14 & 393$\pm$ 47 & 939 $\pm$ 72 \\ \hline
\end{tabular}
\caption{List of participant eccentricity $\varepsilon_{part}$, number
of participants $N_{part}$, and number of binary collisions $N_{bin}$,
from a Glauber calculation~\cite{ms03,glauber} for minimum bias and
three other centrality bins. The errors are statistical from the
calculations only. The biggest systematic error is probably from the
centrality binning based on the impact parameter compared to the
binning of the data based on the multiplicity. All parameters for 62.4
GeV Au+Au collisions are calculated in a similar fashion as for 200
GeV collisions. The p+p cross sections used were 36 mb at 62.4 GeV and
42 mb at 200 GeV.}
\label{tab:glauber}
\end{table}

\begin{figure}[ht]
\center{\includegraphics[width=0.5\textwidth]{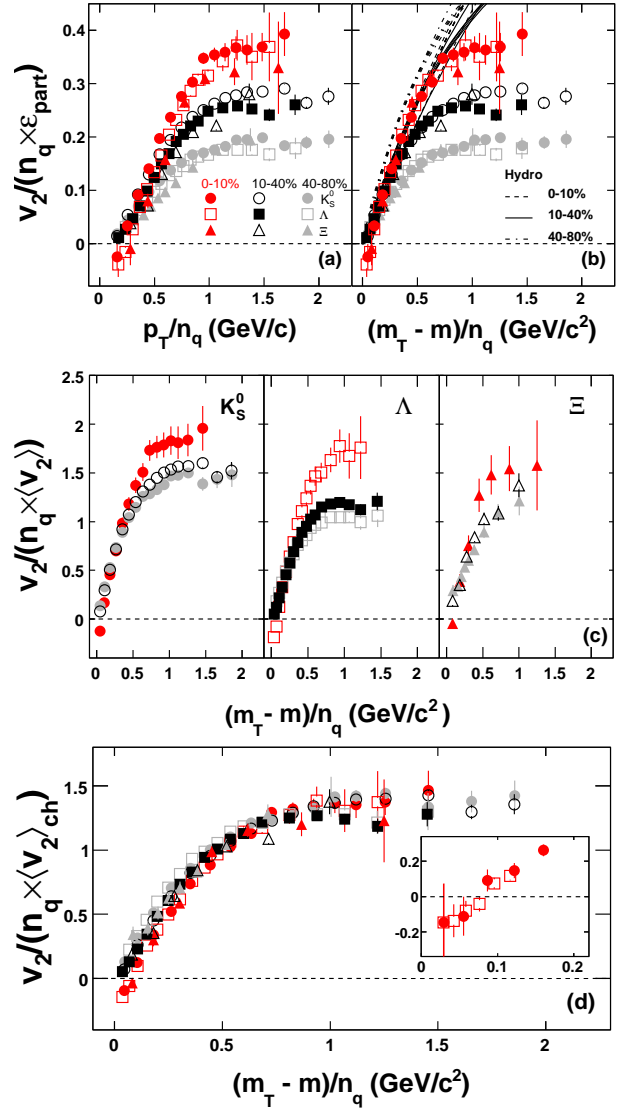}}
\caption{(Color online) $v_2$ scaled by the number of quarks ($n_q$)
and participant eccentricity ($\varepsilon_{part}$), ($v_2/$ ($n_q
\times \varepsilon_{part}$)), of identified particles (particle +
anti-particle) versus (a) the scaled $p_T/n_q$ and (b) ($m_T-m$)/$n_q$
for three centrality bins. For comparison, ideal hydrodynamic model
calculations \cite{pasirev06} are shown as lines in (b). In (c) is
shown the data from (b) scaled by the integrated $v_2$ of each
particle, instead of $\varepsilon_{part}$. In (d) is shown the data
from (b) scaled by the integrated $v_2$ of all charged hadrons. The
insert in (d) expands the low $m_T$ region. The error bars are shown
only for the statistical uncertainties. All data are from
\sqrtsNN = 200 GeV \auau collisions.} \label{ncqeccv2}
\end{figure}

\subsection{Universal scaling?}

In order to analyze the centrality dependence of the scaling
properties, we normalize the $n_q$-scaled elliptic flow $v_2$ by the
participant eccentricity $\varepsilon_{part}$ from a Monte Carlo
Glauber calculation~\cite{ms03,glauber}. (See Table~\ref{tab:glauber}
for $\varepsilon_{part}$.) The results are depicted in
Fig.~\ref{ncqeccv2}. The plots (a) and (b) show the doubly scaled
quantities from three centrality bins as a function of $p_T/n_q$ and
$(m_T-m)/n_q$, respectively. Both plots show an initial rise and a
turn over to a flat region in the higher \pt region. It is interesting
to see in (a) and (b) that at a given centrality, the elliptic flow of
all hadrons are scaled as observed in the minimum bias case
(Fig.~\ref{ncqv2}). After the geometric effect has been removed by
dividing by $\varepsilon_{part}$ in (a) and (b), the build up of
stronger collective motion in more central collisions becomes obvious
in the measured elliptic flow. This is consistent with the ideal
hydrodynamic model calculations, shown as lines in (b), although the
model results are much closer together. However, clearly there is no
scaling amongst different collision centralities. Neither our data nor
the model results indicate universal scaling with eccentricity. A
careful inspection of the results presented in
Ref.~\cite{phenix_scalv2} shows there is no disagreement between data;
rather, the statement of the universal scaling in
Ref.~\cite{phenix_scalv2} is not supported by the data.

To further clarify the issue, instead of dividing the measured $v_2$
by the corresponding eccentricity $\varepsilon_{part}$, we plot
$v_2(m_T-m) / (n_q \times \langle v_2 \rangle)$ for \ks, \lam, and
$\Xi$ in Fig.~\ref{ncqeccv2} (c). The values of $\langle v_2\rangle$
(see Table~\ref{tab:meanv2}) are obtained by averaging $v_2$ as a
function of transverse momentum weighted with the measured spectra. As
one can see in the figure, for a given hadron, this scaling seems to
work better. However, different hadrons seem to have different values
of $v_2$, especially for the top 10\% centrality bin at the higher
$m_T$.

Figure~\ref{ncqeccv2} (d) shows the doubly scaled $v_2$ again. But
this time, the integrated values of $v_2$ are extracted from the
measurements of unidentified charged hadrons $\langle v_2
\rangle_{ch}$ at the corresponding centrality bins. In this case, it
appears that the scaling works better. It is interesting to point out
that at the most central bin, see inset in Fig.~\ref{ncqeccv2} (d),
the values of $v_2$ become negative at low $p_T$ for all hadrons. This
is most likely caused by the strong radial flow developed in central
Au+Au collisions~\cite{pasi01}. Similar behavior has also been
observed in $v_2$ of $\Lambda$ at SPS~\cite{na49lv2}.

\subsection{Integrated $v_2/\epsilon_{part}$ versus collision centrality}

\begin{table}[ht]
\centering
\begin{tabular}{|c||c|c|c|} \hline
  $\sigma_{trig}/\sigma_{geom}$ & 40--80\%  & 10--40\%  & 0--10\%   \\ \hline \hline
  $h^{\pm}$            & 0.0735 $\pm$ 0.000163 & 0.0576 $\pm$ 0.000064& 0.0283 $\pm$ 0.000112\\\hline
 $K_{S}^{0}$              & 0.0707 $\pm$ 0.0008 $\pm$ 0.0013 & 0.0513 $\pm$ 0.0005 $\pm$ 0.0013 & 0.0212 $\pm$ 0.0011 $\pm$ 0.0012 \\ \hline
 $\phi$              & 0.0851 $\pm$ 0.0111 $\pm$ 0.0020 & 0.0658 $\pm$ 0.0082 $\pm$ 0.0016 & 0.0210 $\pm$ 0.0116 $\pm$ 0.0050 (0--5\%) \\ \hline
 $\Lambda$           & 0.0899 $\pm$ 0.0010 $\pm$ 0.0013 & 0.0609 $\pm$ 0.0006 $\pm$ 0.0019 & 0.0221 $\pm$ 0.0012 $\pm$ 0.0029         \\ \hline
 $\Xi$               & 0.0858 $\pm$ 0.0045 $\pm$ 0.0000 & 0.0577 $\pm$ 0.0032 $\pm$ 0.0023 & 0.0220 $\pm$ 0.0028 $\pm$ 0.0015         \\ \hline

\end{tabular}
\caption{The $p_{T}$-averaged $v_{2}$ of identified particles
(particle + anti-particle) from three centrality bins in \sqrtsNN =
200 GeV Au+Au collisions. The Event-Plane method was used to extract
the values of $v_2$. Statistical and systematic errors are shown as
the first and second errors, respectively.}
\label{tab:meanv2}
\end{table}

The integrated elliptic flow values in Table~\ref{tab:meanv2} were
obtained from the measured $v_2(p_T)$ and separately parameterized
$p_T$ spectra. The $v_{2}(p_T)$ were integrated over $p_T$ weighted
with the yield distribution from functions fitted to the spectra. To
extend $v_2$ to low $p_T$, a sixth order polynomial and a
$n_q$-inspired function~\cite{xinv2} were used to fit $v_2$. The $v_2$
values in the table are the average values from these two sets of
parameterizations. The systematic uncertainties are taken as half of
the differences between values from two sets of fits. The statistical
errors as a function of $p_T$ are fitted with third order polynomials
and folded with the yield distributions into the errors of the
integrated $v_2$. The spectra for $K_S^0$ and $\Lambda$ are from
Ref.~\cite{klspectra} and the spectra for $\Xi$ are from
Ref.~\cite{starompt}. Data for $\phi$-mesons are from
Ref.~\cite{star_fv2}.

\begin{figure}[ht]
\center{\includegraphics[width=0.455\textwidth]{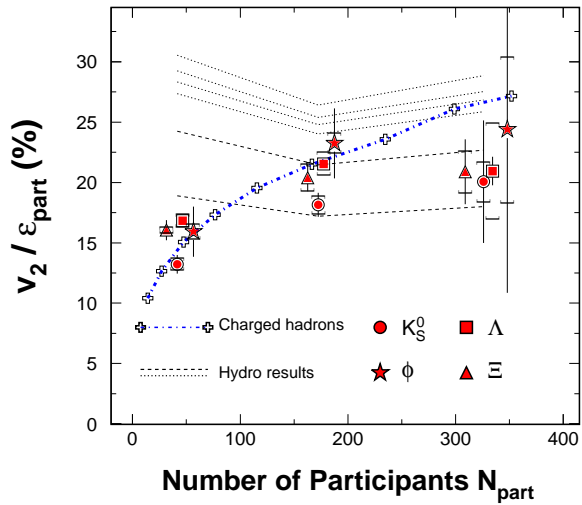}}
\caption{(Color online) Centrality dependence of $v_2 /
\varepsilon_{part}$ versus number of participants for charged
hadrons (crosses), $K_S^0$ (circles), $\phi$
(stars)~\cite{star_fv2}, $\Lambda$ + \alam (squares) and $\Xi$ +
\axi (triangles).  Both unidentified charged
hadron and identified hadron $v_2$ were analyzed with the standard
Event-Plane method. All data are from \sqrtsNN = 200 GeV \auau
collisions. The data points are displaced slightly horizontally for
clarity. The statistical uncertainties and the systematic
uncertainties are shown as bars and brackets, respectively. Ideal
hydrodynamic model calculations are also shown as dashed
lines~\cite{pass06,pasirev06} for, from top to bottom, $\Omega$,
$\Xi$, $\Lambda$, p, K, and $\pi$. } \label{cenv2}
\end{figure}

The centrality dependence of the ratio of the integrated elliptic
flow (Table~\ref{tab:meanv2}) over the eccentricity ($v_2 /
\varepsilon_{part}$) for charged hadrons, \ks,
$\phi$-meson~\cite{star_fv2}, \lam and $\Xi$ are shown in
Fig.~\ref{cenv2}. All these results are from the Event-Plane method
and the number of participants is the average in the centrality
bin. For comparison, results from an ideal hydrodynamic
calculation~\cite{pass06,pasirev06} are also shown as dashed
lines. This ratio, to some extent reflects the strength of the
collective expansion. At more central collision, one would expect a
stronger expansion, hence the larger value of the ratio.  This is what
one sees in Fig.~\ref{cenv2} for charged hadrons. For identified
hadrons, the increasing trend as a function of $N_{part}$ is there
despite the large error bars. In the ideal hydrodynamic
calculations~\cite{pass06,pasirev06}, the first order phase transition
and freeze-out temperatures are set to be 165 MeV and 130 MeV,
respectively. With these parameters, the ideal hydrodynamical model
results describe the pion, kaon and proton transverse momentum
spectra~\cite{pass06,pasirev06}. In a pure hydrodynamic model, one
deals with energy-momentum cells rather than any specific type of
hadrons, thus the initial condition, the equation of state and the
freeze-out conditions used in the calculation are the same for all
hadrons. Such assumptions may not be applicable to all hadrons since
some of them will continue to interact even after hydrodynamic
freeze-out~\cite{hirano06}.

As expected in an equilibrium scenario, the model results show little
sensitivity to the collision centrality. However, it is interesting to
note that there is a clear hadron mass dependence of $v_2$ normalized
by $\varepsilon_{part}$ from the model calculations which is not seen
in the data. It is not clear whether the mass dependence is from the
collective motion at early time or is the effect of the hadronization
process in the calculation. On the data side, the errors are too large
to allow comparisons with model results. As one can see in
Fig.~\ref{cenv2}, after $N_{part} \sim$ 170, the measured ratios for
the strange particles approach that from the ideal hydrodynamic model
calculations. The consistency between model results and data indicates
that the system created in 200 GeV \auau collisions may reach local
thermalization in central collisions when the number of participants
is larger than $\sim$ 170.

\subsection{Energy dependence}

\begin{figure}[ht]
\center{\includegraphics[width=0.7\textwidth]{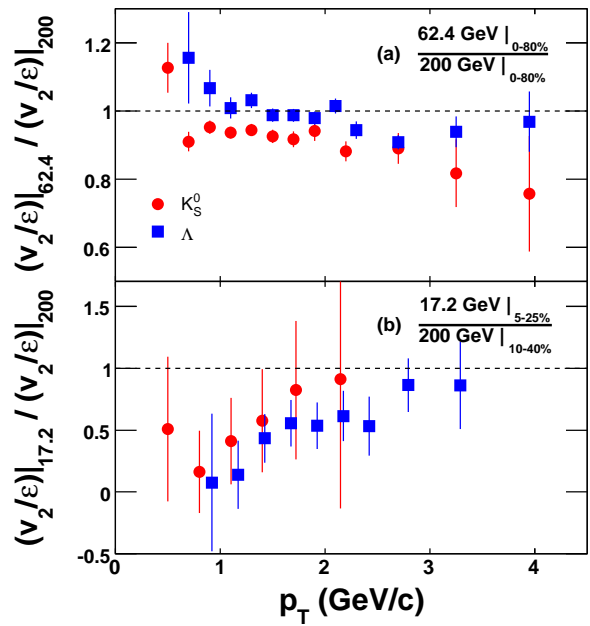}}
\caption{(Color online) The $p_{T}$ dependence of the
eccentricity-scaled $v_{2}$ ratios of (a) 62.4 GeV and (b) 17.2
GeV~\cite{na49lv2} over 200 GeV data for $K_S^0$ (circles) and
$\Lambda$ + \alam (squares). As indicated in (b), the SPS data
points are from 5--25\% and the RHIC data points are from 10--40\%.
Error bars are statistical only.} \label{v2energy}
\end{figure}

The transverse momentum dependence of eccentricity-scaled $v_2$ ratios
are shown in Fig.~\ref{v2energy}. Plots (a) and (b) show the ratio of
data from 62.4 GeV (\auau collisions) and 17.2 GeV ($Pb+Pb$
collisions~\cite{na49lv2}) over 200 GeV \auau collision data,
respectively. It has been observed~\cite{phobos_wp} that since the
charged multiplicity production per participant is proportional to the
square root of the CM energy~\cite{phobos_wp}, $dN/d\eta/\langle
N_{part}/2 \rangle \propto \sqrt{s_{NN}}$, a stronger flow is expected
from higher energy collisions~\cite{voloshin00}. In
Fig.~\ref{v2energy}, the ratios for both $K_S^0$ and $\Lambda$ are
similar. While the higher energy ratios show a decreasing trend as a
function $p_T$, the lower energy ratios seem to increase with
transverse momentum.  In the low $p_T$ region ($\le 1.5$ \GeVc), the
strength of flow is similar in 200 GeV and 62.4 GeV
\auau collisions. For the 0--80\% Au+Au collisions, the values of the
participant eccentricity from 62.4 and 200 GeV are 0.392 and 0.384,
respectively.  The lack of energy dependence in $v_2$ for $K_S^0$ and
$\Lambda$ is due to the similarity in the participant
eccentricity. The PHOBOS experiment reported a similar observation for
the $v_2$ of charged hadrons~\cite{phobos_prl}. As discussed
in~\cite{phobos_wp}, from 62 GeV to 200 GeV, the
participant-normalized charged hadron density at mid-rapidity
increased by about 50\%. However, we do not observe a change of a
similar size in the participant eccentricity and $v_2$, indicating
that a large fraction of the particle production occurs at the later
stage of heavy ion collisions at these beam energies.

The ratio is less than unity in the low $p_T$ region (see
Fig.~\ref{v2energy} (b)) indicating that the flow is weaker in the
lower energy $Pb+Pb$ collisions~\cite{na49lv2}. The collective
velocity parameters, extracted from the transverse momentum spectra,
are also found to be larger in 200 GeV \auau collisions than those from
17.2 GeV $Pb+Pb$ collisions~\cite{starwp}. Since elliptic flow
develops at a relatively early stage of the collisions, the observed
increase in collective flow in \auau collisions at RHIC is therefore
caused by early partonic
interactions~\cite{starv21,PIDv2130,flowPRC}.

\begin{figure}[ht]
\center{\includegraphics[width=0.7\textwidth]{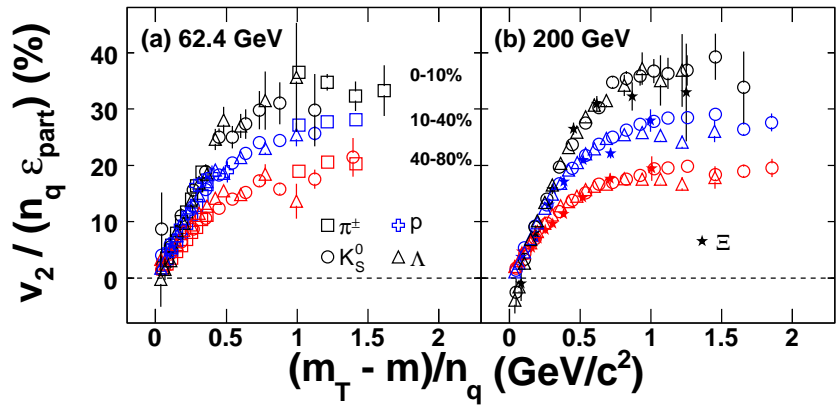}}
\vspace{-1.0cm} \caption{(Color online) Centrality dependence of
$v_{2}/(n_q \varepsilon_{part})$ for identified hadrons (particle +
anti-particle) from (a) 62.4 GeV~\cite{star_v262} and (b) 200 GeV
\auau collisions. Error bars are statistical only. For the
10--40\% centrality the systematic uncertainty should be the same as
in Fig.~\ref{mainv2pt}.}
\label{v2energy2}
\end{figure}

Figure~\ref{v2energy2} shows the centrality dependence of $v_2$
normalized by the number of quarks and eccentricity ($v_{2}/(n_q
\varepsilon_{part})$) for identified hadrons from (a) 62.4
GeV~\cite{star_v262} and (b) 200 GeV \auau collisions. Within error
bars, data from both energies are similar. At the low transverse
energy region, the scaled $v_2$ shows almost a linear increase and
then becomes flat.  For more central collisions the turning point is
at higher values of $(m_T-m)/n_q$. Recently, PHENIX has reported a
charged hadron scaling with eccentricity, system size, and the
transverse energy $(m_T-m)$ up to 1~GeV/$c^2$~\cite{phenix_scalv2}.
As one can see from the figure, at a given centrality, independent of
the collision energy, there is a clear scaling: all values of
$v_2/(n_q \varepsilon_{part})$ coalesce into a single distribution.
On the other hand, it is clear in the figure that at different
centralities the shape of the distributions are different, meaning
that there is no scaling in the measured $v_2$ with the eccentricity,
especially in the higher transverse energy region.

\section{Summary}
\label{concl}

We present STAR results on the elliptic flow $v_2$ of unidentified
charged hadrons, strange and multi-strange hadrons from \sqrtsNN~= 200
GeV \auau collisions at RHIC. The centrality dependence of $v_2$ over
a broad transverse momentum range is presented. Comparison of
different analysis methods are made in order to estimate systematic
uncertainties. The rapidity dependence of the charged hadron $v_2$
from these measurements is consistent, at both mid-pseudorapidity and
forward rapidities, with both STAR~\cite{flowPRC} and
PHOBOS~\cite{phobos} reported results. In particular, the results for
$v_2$ from the Lee-Yang Zero method for charged hadrons, \ks, and \lam
are shown for the first time at RHIC. The non-flow effects, studied in
the 10--40\% centrality window, are on the order of 10\% within 0.2
$\le p_T \le 3$ \GeVc\: and up to 25\% at $p_T \sim 6$ \GeVc.

In the relatively low \pt region, \pt $\le$ 2 \GeVc, a scaling with
$m_T - m$ is observed for identified hadrons under study in each
centrality bin and there is a clear centrality dependence in the
scaling. However, we do not observe $v_2(m_T-m)$ scaled by the
participant eccentricity to be independent of centrality. The largest
values of the participant eccentricity scaled $v_2$ are in the most
central collisions. For the most central collisions (0--10\%),
negative values of $v_2$ at the lowest \pt studied have been observed
for both \ks and \lam. This is the first time a negative $v_2$ in
\auau collisions at RHIC has been found. It is consistent with
the strong expansion observed in hadron spectra
analysis~\cite{starwp}. In the higher \pt region, 2 $\le$ \pt $\le$ 6
\GeVc, number-of-quark scaling is observed for all particles under
study. For the multi-strange hadron $\Omega$, which does not suffer
appreciable hadronic interactions, the values of $v_2$ are consistent
with both $m_T -m$ scaling at low \pt and number-of-quark scaling at
intermediate $p_T$.

As a function of collision centrality, an increase of $p_T$-integrated
$v_2$ scaled by the participant eccentricity has been observed,
indicating stronger collective flow in more central \auau
collisions. However, in the higher transverse energy region there is
no scaling of $v_2$ with eccentricity.

The energy dependence of $v_2$ for \ks and \lam was presented for
\sqrtsNN~= 17 GeV $Pb + Pb$, and 62 and 200 GeV \auau collisions. No clear
systematic trend was observed, but the differences were only on the
order of 10\%.

For comparison, results from ideal hydrodynamical model calculations
were used.  The calculations over-predict the data at $p_T \ge$ 2
\GeVc. At low \pt, although the model predicts correctly the mass
hierarchy observed in the data, there is no scaling with $m_T-m$, and
no scaling with the number of quarks is observed throughout the \pt
region for all hadrons. We observe that the mass ordering at low \pt
alone is not sufficient to claim thermalization in \auau collisions at
RHIC.

\section{Acknowledgments}

We thank Jean-Yves Ollitrault and Nicolas Borghini for help in
understanding the Lee-Yang Zero method.

We thank the RHIC Operations Group and RCF at BNL, and the NERSC
Center at LBNL and the resources provided by the Open Science Grid
consortium for their support. This work was supported in part by the
Offices of NP and HEP within the U.S. DOE Office of Science, the
U.S. NSF, the Sloan Foundation, the BMBF of Germany, CNRS/IN2P3, RA,
RPL, and EMN of France, EPSRC of the United Kingdom, FAPESP of Brazil,
the Russian Ministry of Sci. and Tech., the NNSFC, CAS, MoST and MoE
of China, IRP and GA of the Czech Republic, FOM of the Netherlands,
DAE, DST, and CSIR of the Government of India, Swiss NSF, the Polish
State Committee for Scientific Research, Slovak Research and
Development Agency, and the Korea Sci. and Eng.  Foundation.


\clearpage

\end{document}